\documentclass[preprint,showpacs,preprintnumbers,amsmath,amssymb]{revtex4}
\usepackage{graphicx}

\begin{document}

\thispagestyle{empty}

\title{Possibility to measure thermal effects in the
Casimir force}

\author{B.~Geyer,${}^{1}$
G.~L.~Klimchitskaya,${}^{1,2}$ and
V.~M.~Mostepanenko${}^{1,3}$}

\affiliation{${}^1$Institute for Theoretical
Physics, Leipzig University, Postfach 100920,
D-04009, Leipzig, Germany \\
${}^2${North-West Technical University,
Millionnaya Street 5, St.Petersburg,
191065, Russia}\\
${}^3${Noncommercial Partnership ``Scientific Instruments'',
Tverskaya Street 11, Moscow,
103905, Russia}
}

\begin{abstract}
We analyze the possibility to measure small thermal effects in
the Casimir force between metal test bodies in configurations
of a sphere above a plate and two parallel plates. For sphere-plate
geometry used in many experiments we investigate the applicability
of the proximity force approximation (PFA) to calculate thermal
effects in the Casimir force and its gradient. It is shown that
for real metals the two formulations of the PFA used in the
literature lead to relative differences in the obtained results
being less than a small parameter equal to the ratio of separation
distance to sphere radius. For ideal metals the PFA results for the
thermal correction are obtained and compared with available exact
results. It is emphasized that in the experimental region in the
zeroth order of the small parameter mentioned above the thermal
Casimir force and its gradient calculated using the PFA (and
thermal corrections in their own right) coincide with respective
exact results. For real metals available exact results are outside
the application region of the PFA. However, the exact results are
shown to converge to the PFA results when the small parameter goes
down to the experimental values. We arrive at the conclusion that
large thermal effects predicted by the Drude model approach, if
existing at all, could be measured in both static and dynamic
experiments in sphere-plate and plate-plate configurations. As to
the small thermal effects predicted by the plasma model approach,
the static experiment in the configuration of two parallel plates
is found to be the best for its observation.
\end{abstract}
\maketitle

\section{Introduction}

The Casimir effect \cite{1} is now universally known as one of the
most extensively studied manifestations of zero-point oscillations.
It attracts considerable attention in quantum field theory,
gravitation and cosmology, atomic physics and optics, condensed
matter physics, and in applications to nanotechnology
(see monographs \cite{2,3,4,5,6}).
The fundamental theory describing both the van der Waals and
Casimir forces between two semispaces with planar boundaries was
developed by Lifshitz \cite{7,8}. At present the Lifshitz theory is
generalized for material bodies with curved boundary surfaces
(see, e.g., Refs.~\cite{9,10}). Over a long period of years only a very
limited experimental information about the Casimir force was
available. Recently, however, a significant advance has been made
in the experimental study of this phenomenon reflected in
review \cite{11}.
In principle, the theoretical and experimental progress taken together
allows computation and measurement of Casimir forces between bodies of
complicated geometrical shape made of different materials.
In order for such computation to be made, one should know the reflection
amplitudes on the boundary surfaces at all Matsubara frequencies.
At first sight this should present no problems because the reflection
amplitudes can be either measured directly or calculated using
measured material properties, such as frequency-dependent dielectric
permittivities. At this point, however, difficulties emerge unexpectedly
which are connected with the role of relaxation properties of charge
carriers in the Casimir effect.

Beginning in 2000, the influence of relaxation on the thermal Casimir
force was hotly debated. Bostr\"{o}m and Sernelius \cite{12} have
noticed that the substitution of the dielectric permittivity of the
Drude model with nonzero relaxation parameter into the Lifshitz
formula results in a decreasing magnitude of the Casimir
free energy and
force as a function of temperature over a wide region of
separations. This is in contradiction with the case of ideal metal
plates or plate materials described by the nondissipative plasma  model,
where the magnitudes  of the Casimir free energy and  force
are monotonously increasing functions of the temperature \cite{12a,12b}.
At large separations (or in high temperature limit) the magnitudes
of the Casimir
free energy and  force between metal plates calculated using
the Drude model are by a factor of 2 less then the same quantities
calculated for ideal metals or metals described by the plasma model.
It was also shown that for metals with perfect crystal lattices the
Lifshitz theory combined with the Drude model violates the Nernst heat
theorem \cite{6,11,13,14}. For metals with impurities leading to a
nonzero relaxation at zero temperature Nernst's theorem was shown to be
followed \cite{15,16}. This, however, does not solve the problem because
perfect crystal metal is an idealized model of a truly equilibrium system
with nondegenerate ground state for which the laws of thermodynamics
must be satisfied.

A large thermal effect in the Casimir force between metals at separations
of a few hundred nanometers predicted by the Lifshitz theory combined
with the Drude model was experimentally excluded \cite{17} by indirect
dynamic measurements of the Casimir pressure in the configuration of
a sphere above a plate of a micromachined oscillator.
Later this experiment was repeated for two more times with increased
precision. The exclusion of the thermal effect predicted by the Drude
model was confirmed at a 95\% confidence level \cite{18,19} and at
a 99.9\% confidence level \cite{20,21}. The same measurement data
were found to be consistent with the Lifshitz theory combined with the
plasma model. It is important to keep in mind that the comparison of
experiment with theory in Refs.~\cite{17,18,19,20,21} was based on the
use of an approximate method, the so-called {\it proximity force
approximation} (PFA) \cite{6,11,22} because
at that time for sphere-plate
configuration an exact theory was not available. At the moment there is
a conceptual possibility to compute
the thermal Casimir force between a
sphere and a plate made of real metals with no use of the PFA \cite{23,24},
but the region of experimental parameters is not yet achieved due to
computational difficulties.

When it is considered that the Drude model correctly describes
the relaxation
of conduction electrons at low frequencies, the contradiction with
basic laws of thermodynamics and disagreement with the experimental
data outlined above are puzzling. These problems were dramatized by the
demonstration that the inclusion of dc conductivity of dielectric (or
dielectric-type semiconductor) materials into the model of dielectric
response in the Lifshitz theory also results in a violation of the Nernst
theorem \cite{25,26,27,28}. From the experimental side, it was shown that
the measurement data of the experiment on optical modulation of the
Casimir force between Au sphere and Si plate with light \cite{29,30}
excludes the Lifshitz theory taking into account the dc conductivity of
dielectrics at a 95\% confidence level.
A similar result was obtained from the measurement of the Casimir-Polder
force between the Bose-Einstein condensate of ${}^{87}$Rb atoms and
SiO${}_2$ plate \cite{31}. Here, the Lifshitz theory taking dc
conductivity of SiO${}_2$ into account was experimentally excluded at
a 70\% confidence level \cite{32}. It is pertinent to note that while the
comparison of the optical modulation experiment with theory uses the
PFA, the measurement results for the Casimir-Polder force were compared
with the exact Lifshitz formula for atom-wall interaction.
One can summarize that experiments with metals, semiconductors and
dielectrics exclude the influence of dissipation of conduction
electrons on the Casimir force (the statement on the opposite in
the Introduction to Ref.~\cite{24} is a typo).

The conflict between such a fundamental theory, as the Lifshitz theory,
thermodynamics and the experimental data of several experiments is a
problem of great concern. Because of this, a lot of attempts to
resolve this problem has been
undertaken. Specifically, it was even suggested
\cite{33,34} to modify the Lifshitz theory by including into
consideration screening effects and diffusion currents. It was noted,
however, that the modifications proposed do not alleviate contradictions
with thermodynamics and the experimental data (discussion on this
subject can be found in Refs.~\cite{35,36,37,38,39,40}).
On the other hand, it was suggested \cite{41} to modify the Planck
distribution law by taking into account  ``saturation effects''.
There were also attempts to soften contradictions with thermodynamics
by reformulating the problem \cite{42} and by finding additional
statistical arguments in favor of the Drude model \cite{43,44}.
In Ref.~\cite{46aa}, in addition to the usually used exponential
screening, the so-called {\it algebraic screening} in atom-wall
interaction was considered. As a result, linear in temperature
thermal correction to the Casimir-Polder force at short
separations was predicted similar to that predicted by the
Drude model. Note that the algebraic screening is connected with
nonanalytic terms in the small wavenumber expansion of the
dielectric permittivity.
It was finally suggested \cite{45} that for the resolution of the
problem some concepts of statistical physics related to the theoretical
description of the interaction of classical and quantum fluctuating
fields with matter might need a reconsideration.

The prospects for a pure theoretical resolution of the above problems
seem dim at the moment. In this situation any additional experimental
evidence could be very useful. In this paper we analyze the possibility
to measure thermal effects in the Casimir force on the basis of already
created and used experimental setups. We stress that in the experiments
\cite{17,18,19,20,21,29,30} mentioned above the large thermal
correction to the Casimir force, as predicted by the Drude model,
was excluded. However, these experiments were not of sufficient
precision to measure the thermal effect for metals predicted by the
plasma model or for dielectrics with dc conductivity omitted
(till the moment the thermal effect was measured in the Casimir-Polder
force alone \cite{31}). Keeping in mind that there is some confusion
in the literature concerning the use of the PFA, we present two
(not equivalent) formulations of the PFA and clarify which of them was
really used in the comparison between experiment and theory.
We especially analyze the calculation results for the thermal correction
to the Casimir force in sphere-plate configuration found using the PFA
and explain when they are meaningful. The obtained conclusions are
confirmed by the comparison with exact results for the thermal
contribution to the Casimir force between a sphere and a plate.
We show that at the moment it is not possible to measure small thermal
effects in the Casimir force (or its gradient) in sphere-plate
geometry, as predicted by the Lifshitz formula combined with the plasma
model. We also discuss the configuration of two parallel plates
in both dynamic and static regimes. According to our results, small
thermal corrections to the gradient of the Casimir pressure in the
dynamic regime is suppressed. The only way to measure small thermal
effects in the Casimir pressure is suggested by the configuration
of two parallel plates in the static regime.

The paper is organized as follows. In Sec.~II two different formulations
of the PFA are considered and applied to the Casimir force in
sphere-plate configuration. Section~III discusses the same subject with
respect to the gradient of the Casimir force between a sphere and
a plate. In Sec.~IV the relationship between the PFA and the exact
results is presented for a sphere and a plate made of ideal metal.
The applicability of the PFA to describe thermal corrections to the
Casimir force between a sphere and a plate made of real metals
and the possibility to measure the thermal effect are
considered in Sec.~V. In Sec.~VI we show that the static Casimir
configuration of two parallel plates is preferential for the
observation of a small thermal effect in the  Casimir pressure.
Section~VII contains our conclusions and
discussion.

\section{The proximity force approximation for the Casimir
force between a sphere and a plate}

Different authors vary somewhat in the meaning of the term ``PFA''.
In fact the term ``proximity force theorem'' (later changed for PFA)
was introduced in Ref.~\cite{22} where the so-called
{\it Derjaguin method} \cite{46} was applied in order to calculate
the force acting between curved surfaces by using the known force
per unit area of plane parallel plates. In the Derjaguin method,
the unknown force between the elements of curved surfaces is
approximately replaced with a known force per unit area of plane
surfaces at respective separations. In application to a sphere of
radius $R$ above a plane surface of a plate $z=0$ the Derjaguin
method represents the force between them in the form
\begin{equation}
F_{\rm sp}(a,T)=\int_{\Sigma}d\sigma\,P(z,T).
\label{eq1}
\end{equation}
\noindent
Here, $d\sigma$ is the element of plate area, $\Sigma$ is the
projection of the sphere onto the plate, $a$ is the shortest separation
between the sphere and the plate, and $P(z,T)$ is the force per unit area
of two plane parallel plates at a separation $z$ at temperature $T$
(i.e., the Casimir pressure). Choosing the origin of a cylindrical
coordinate system on the plane $z=0$ under the sphere center,
the coordinate $z$ of any point on the sphere is given by
$z=R+a-(R^2-\rho^2)^{1/2}$.
Then Eq.~(\ref{eq1}) leads to
\begin{eqnarray}
F_{\rm sp}(a,T)&=&2\pi\int_{0}^{R}\rho\,d\rho\,P(z,T)
\label{eq2} \\
&=&
2\pi\int_{a}^{R+a}(R+a-z)P(z,T)\,dz.
\nonumber
\end{eqnarray}
\noindent
Keeping in mind that the thermal Casimir pressure is connected
with the free energy per unit area of two parallel plates as
\begin{equation}
P(z,T)=-\frac{\partial{\cal F}_{\rm pp}(z,T)}{\partial z},
\label{eq3}
\end{equation}
\noindent
and integrating by parts in Eq.~(\ref{eq2}), one arrives at
\begin{equation}
F_{\rm sp}(a,T)=2\pi R{\cal F}_{\rm pp}(a,T)
-2\pi\int_{a}^{R+a}dz\,{\cal F}_{\rm pp}(z,T).
\label{eq4}
\end{equation}
\noindent
This generalizes Eq.~(20) of Ref.~\cite{46a} related to the
nonretarded case.

Further simplification of Eq.~(\ref{eq4}) can be achieved when it is
assumed that the free energy ${\cal F}_{\rm pp}(z,T)$ is a quickly
decreasing function of $z$ and drops to zero on the characteristic
length of about the sphere radius $R$. Let us consider the case of
the free energy of the Casimir interaction given by the Lifshitz
formula,
\begin{equation}
{\cal F}_{\rm pp}(z,T)=\frac{k_B T}{2\pi}
\sum_{l=0}^{\infty}{\vphantom{\sum}}^{\prime}
\int_{0}^{\infty}k_{\bot}dk_{\bot}
\sum_{\alpha}\ln\left(1-r_{\alpha}^2e^{-2q_lz}\right).
\label{eq5}
\end{equation}
\noindent
Here, $k_B$ is the Boltzmann constant, the prime near the summation
sign multiplies the term with $l=0$ by 1/2, $k_{\bot}$ is the
projection of the wave vector on the plane of plates, and
$q_l=(k_{\bot}^2+\xi_l^2/c^2)^{1/2}$ where
$\xi_l=2\pi k_B Tl/\hbar$ with $l=0,\,1,\,2,\,\ldots$ are the
Matsubara frequencies. The reflection coefficients $r_{\alpha}$ for the
two polarizations of the electromagnetic field, transverse magnetic
($\alpha={\rm TM}$) and transverse electric
($\alpha={\rm TE}$), are expressed in terms of the dielectric permittivity
along the imaginary frequencies,
\begin{eqnarray}
&&
r_{\rm TM}=r_{\rm TM}(i\xi_l,k_{\bot})=
\frac{\varepsilon(i\xi_l)q_l-k_l}{\varepsilon(i\xi_l)q_l+k_l},
\nonumber \\
&&
r_{\rm TE}=r_{\rm TE}(i\xi_l,k_{\bot})=
\frac{q_l-k_l}{q_l+k_l},
\label{eq6} \\
&&k_l=\left[k_{\bot}^2+\varepsilon(i\xi_l)
\frac{\xi_l^2}{c^2}\right]^{1/2}.
\nonumber
\end{eqnarray}
\noindent
In order to simplify Eq.~(\ref{eq4}), we use an expansion in power
series in Eq.~(\ref{eq5}):
\begin{equation}
{\cal F}_{\rm pp}(z,T)=-\frac{k_B T}{2\pi}
\sum_{l=0}^{\infty}{\vphantom{\sum}}^{\prime}
\sum_{n=1}^{\infty}\frac{1}{n}
\int_{0}^{\infty}k_{\bot}dk_{\bot}
\sum_{\alpha}r_{\alpha}^{2n}e^{-2q_lnz}.
\label{eq7}
\end{equation}
\noindent
Then from Eq.~(\ref{eq7}) one finds
\begin{eqnarray}
&&
I(a,T)\equiv -\int_{a}^{R+a}dz{\cal F}_{\rm pp}(z,T)=\frac{k_B T}{4\pi}
\sum_{l=0}^{\infty}{\vphantom{\sum}}^{\prime}
\sum_{n=1}^{\infty}\frac{1}{n^2}
\nonumber \\
&&~~~~\times
\int_{0}^{\infty}\frac{k_{\bot}dk_{\bot}}{q_l}
\sum_{\alpha}r_{\alpha}^{2n}\left[e^{-2q_lna}-
e^{-2q_ln(R+a)}\right].
\label{eq8}
\end{eqnarray}
\noindent
When $R$ goes to infinity with $a$ and $T$ fixed, the contribution
to $I(a,T)$
of the second term in square brackets on the right-hand side of
Eq.~(\ref{eq8})  vanishes as $(R+a)^{-1}$.
This means that for large $R$ the value of $I(a,T)$ is determined by
the first term in square brackets and can  be considered as independent
on $R$. Hence, rewriting Eq.~(\ref{eq4}) in the form
\begin{equation}
F_{\rm sp}(a,T)=2\pi R{\cal F}_{\rm pp}(a,T)\left[1+
\frac{I(a,T)}{{R}{\cal F}_{\rm pp}(a,T)}\right],
\label{eq9}
\end{equation}
\noindent
one concludes that in the limit of large $R$ it holds
${I(a,T)}/{{R}{\cal F}_{\rm pp}(a,T)}\sim C/R$ where $C$ is some constant.

Let us now consider the behavior of the quantity $I/R{\cal F}_{\rm pp}$
in the limiting case $a\to 0$ keeping $R$ fixed.
This is a nonrelativistic limit where \cite{6,47}
\begin{equation}
{\cal F}_{\rm pp}(a,T)=E_{\rm pp}(a)=-\frac{H}{12\pi a^2},
\quad
I(a,T)=\frac{RH}{12\pi a(R+a)}
\label{eq10}
\end{equation}
\noindent
with the Hamaker constant defined by
\begin{equation}
H=\frac{3\hbar}{8\pi}\int_{0}^{\infty}d\xi
\int_{0}^{\infty}y^2dy\left\{\left[
\frac{\varepsilon(i\xi)+1}{\varepsilon(i\xi)-1}\right]^2
e^{y}-1\right\}^{-1}.
\label{eq11}
\end{equation}
\noindent
{}From Eq.~(\ref{eq10}) it follows
${I(a,T)}/{{R}{\cal F}_{\rm pp}(a,T)}\sim Ca$ when $a$
vanishes. Thus Eq.~(\ref{eq9}) can be rewritten as
\begin{equation}
F_{\rm sp}(a,T)=2\pi R{\cal F}_{\rm pp}(a,T)\left[1+
f(a,T)\frac{a}{R}\right],
\label{eq12}
\end{equation}
\noindent
where $f(a,T)$ is scarcely affected by the sphere radius $R$.
For all models of dielectric permittivity used to describe metals
the magnitude of $f(a,T)$ is less than unity over wide ranges of
experimental parameters (see below the results of numerical
computations).

Under the condition $a\ll R$ which is usually valid in experiments on
measuring the Casimir force one may neglect the term of order $a/R$
on the right-hand side of Eq.~(\ref{eq12}) and arrives at
\begin{equation}
F_{\rm sp}(a,T)=2\pi R{\cal F}_{\rm pp}(a,T).
\label{eq13}
\end{equation}
\noindent
Just this equation was called the ``proximity force theorem'' in
Ref.~\cite{22} and heavily used in the comparison of experiment
with theory in all measurements of the Casimir force in sphere-plate
geometry (see, e.g., Refs.~\cite{29,30,48,49,50,51,52,53,54,55,56,57,58}).
As can be seen from the above derivation, Eq.~(\ref{eq13}) follows from
Eq.~(\ref{eq1}) under some conditions, but is not equivalent to it.
Because of this, to avoid confusion, Ref.~\cite{59} suggested to call
(\ref{eq1}) the {\it most general formulation} of the PFA and
(\ref{eq13}) the {\it simplified formulation} of the PFA.
Keeping in mind that the used in experiments simplified formulation
is obtained by disregarding contributions of order $a/R$, it would be
meaningless to attribute physical meaning to any terms of order
$a/R$ in the thermal Casimir force $F_{\rm sp}$ calculated by using
Eq.~(\ref{eq13}) (see discussion in Secs.~IV and V).

One further version of the PFA discussed in the literature \cite{60,61}
is connected with the choice of parallel surface elements in Eq.~(\ref{eq1}).
In the original Derjaguin method \cite{46} used by us the surface elements
representing the sphere are parallel to the surface of the plate $z=0$.
In this case $\Sigma$ is a part of the plane $z=0$ (the so-called
{\it plate-based} PFA). If, however, the lower half of
the sphere is chosen as
$\Sigma$ \cite{60}, the plane elements representing the sphere are
tangential to it and respective twin elements of the plate are tilted
by different angles with respect to the plane $z=0$ (the so-called
{\it sphere-based} PFA). In both cases,
the distance between the elements is
measured along the normal to $\Sigma$. It was shown \cite{60} that both
the plate-based and sphere-based PFA lead to coinciding results in the
zeroth order of $a/R$, i.e., to Eq.~(\ref{eq13}). The form of the function
$f(a,T)$ in Eq.~(\ref{eq12}) for both versions of the PFA is, however,
different. Basing on this, Refs.~\cite{61,62} considered the
results for the Casimir energies and forces obtained by PFA
as ambiguous. The differences
between the two versions of the PFA were treated as some ``error bars''
inherent to this approximate method. Keeping in mind, however, that the
comparison of experiment with theory in sphere-plate geometry is based
not on the general formulation of the PFA (\ref{eq1}), but on an
unambiguous simplified formulation (\ref{eq13}), the discussion of
inherent to PFA errors is immaterial. In fact, when speaking about the
Casimir force, only the zeroth order in $a/R$ results in any of the
PFA formulations are of physical significance and only under the
condition $a\ll R$.

The situation changes drastically when, instead to the Casimir force,
the PFA is applied, for instance, to the gravitational force.
According to Refs.~\cite{59,62a}, the most general formulation of the PFA
(\ref{eq1}) leads to an exact result for the force between a sphere
and a plate for all conservative volumetric forces,
particularly for the gravitational force.
 In so doing, however, only the original Derjaguin choice of
$\Sigma$ (plate-based) must be used. This makes the plate-based
version of the PFA preferable in comparison with the sphere-based
version.

Now we present the results of numerical computations for the function
$f(a,T)$ defined in Eq.~(\ref{eq12}) in typical regions of experimental
parameters. Computations were performed by using the Lifshitz formula
(\ref{eq5}) and Eq.~(\ref{eq8}) for a sphere and a plate made of Au.
The dielectric properties of Au were described by using three
different models. As a crude approximation, the model of ideal metal
bodies was used leading to $r_{\rm TM}(i\xi_l,k_{\bot})=1$,
$r_{\rm TE}(i\xi_l,k_{\bot})=-1$. A frequently used description obtains the
dielectric permittivity $\varepsilon(i\xi_l)$ by means of the
Kramers-Kronig relation
\begin{equation}
\varepsilon(i\xi_l)=1+\frac{2}{\pi}\int_{0}^{\infty}
\frac{\omega\,{\rm Im}\,\varepsilon(\omega)}{\omega^2+\xi_l^2}\,d\omega,
\label{eq14}
\end{equation}
\noindent
where ${\rm Im}\,\varepsilon(\omega)$ is taken from tables of the
optical data for Au \cite{63} extrapolated to low freuqencies by means of
the Drude model
\begin{equation}
{\rm Im}\,\varepsilon(\omega)=
\frac{\omega_p^2\gamma}{\omega(\omega^2+\gamma^2)}.
\label{eq15}
\end{equation}
\noindent
Here, the plasma frequency and the relaxation parameter of Au are
given by $\omega_p=9.0\,$eV, $\gamma=0.035\,$eV \cite{64}.
As one more alternative description used in the literature we have
applied the generalized plasma-like model \cite{6,11,21}
\begin{equation}
\varepsilon(i\xi_l)=1+\frac{\omega_p^2}{\xi_l^2}+
\sum_{j=1}^{6}\frac{g_j}{\omega_j^2+\xi_l^2+\gamma_j\xi_l},
\label{eq16}
\end{equation}
\noindent
where $\omega_j\neq 0$ are the resonant frequencies of the oscillators
describing core electrons, $\gamma_j$ are the relaxation
frequencies, and $g_j$ are the oscillator strengths. The values of all
these parameters for Au can be found in \cite{6,21}.
Note that there are proposals in the literature for alternative
dielectric functions taking into account the effect of spatial
nonlocality (see, for instance, Refs.~\cite{40,67a,67b}).
However, for metallic test bodies these dielectric functions
predict precisely the same thermal effect as the Drude model
approach \cite{39,67a,67b}. Because of this, they do not require
a special consideration here.

In Fig.~1 the computational results for $f(a,T)$ are shown as a function
of (a) separation at room temperature $T=300\,$K and (b) temperature at
a separation $a=2\,\mu$m. In both cases a sphere with
$R=100\,\mu$m radius is used. The dotted, dashed and solid lines
demonstrate the results obtained using ideal metals, Au described by the
optical data extrapolated by the Drude model and Au described by the
generalized plasma-like model, respectively. As can be seen in Fig.~1(a),
in the region of separations from 100\,nm to $5\,\mu$m it holds
$0.5\leq|f(a,T)|< 1$. It can be easily verified that in the limiting
case $a\to 0$ the function $f(a,T)\to -1/2$ for ideal metals (the
dotted line) and $f(a,T)\to -1$ for real Au independently of the model
used for its description (the dashed and solid lines). {}From Fig.~1(b)
it is seen that at any temperature from absolute zero to 300\,K the
magnitudes of $f(a,T)$ remain less than unity and the difference
between various models of
dielectric properties disappears with vanishing temperature.

Now we consider the measure of dependence of
the function $f(a,T)$ on the
sphere radius. The computational results for $f(a,T)$ as a function of
$\log_{10}(a/R)$ are shown in Fig.~2 as the three groups of lines (dotted,
dashed and solid for ideal metals and real Au described using the Drude
and plasma models as explained above) numerated 1,\,\,2, and 3 for fixed
separations $a=0.1$, 2, and $5\,\mu$m, respectively. Computations are
done at room temperature $T=300\,$K for $a/R$ varying from $10^{-4}$ to
0.05. For example, as is seen in Fig.~2, at $a=0.1\,\mu$m (the group 1 of
lines) there is almost no dependence of $f(a,T)$ on $R$ for radii
larger than $2\,\mu$m. For $a=2\,\mu$m (the group 2 of
lines) $f(a,T)$ does not depend on $R$ for radii $R>200\,\mu$m.
We remind that typical sphere radii are $R=100\,\mu$m in the experiments
 \cite{29,30,49,50,51,52,53,54,55,56,57,58}
and $R=150\,\mu$m in the experiments
\cite{18,19,20,21}, performed at separations from less than 100\,nm
to a few hundred nanometers. In Fig.~2 it is seen also that the
relative magnitude of the function $f(a,T)$ computed using different
models of dielectric permittivity of Au depends on separation.
For example, at $a=0.1\,\mu$m $|f(a,T)|$ computed using the model of
ideal metals (the dotted line) is less than using the generalized
plasma-like model (the solid line) and using the tabulated optical data
extrapolated by the Drude model (the dashed line). At separations $a=2$ and
$5\,\mu$m the magnitude of $f(a,T)$ computed using the model of ideal
metals is sandwiched between lines computed using the two models of
real Au.

The size of possible corrections to the simplified formulation of the PFA
(\ref{eq13}) has been investigated experimentally by measuring the
Casimir force between an Au-coated plate and five Au-coated spheres with
different radii using a micromachined oscillator \cite{65}.
In so doing spheres with radii $R=10.5$, 31.4, 52.3, 102.8 and
$148.2\,\mu$m were used. The obtained constraint $|f(a,T)|\leq 0.6$ for
$a<300\,$nm, $T=300\,$K is in very good agreement with the
computational results in Fig.~1(a). Thus, the experimental data are in favor
of the simplified formulation of the PFA.

\section{The proximity force approximation for the
gradient of the Casimir
force between a sphere and a plate}

In many experiments using the sphere-plate configuration separation
distance between the test bodies was varied harmonically (see, e.g.,
Refs.~\cite{6,11,17,18,19,20,21}). In this case not the Casimir force
but its gradient is the physical quantity immediately connected
with the frequency shift of a micromachined oscillator. The gradient
of the Casimir force acting between a sphere and a plate can be found
by differentiating Eq.~(\ref{eq4})  with respect to $a$ and taking
into account Eq.~(\ref{eq3}),
\begin{equation}
\frac{\partial F_{\rm sp}(a,T)}{\partial a}=
-2\pi RP(a,T)+2\pi{\cal F}_{\rm pp}(a,T)-2\pi{\cal F}_{\rm pp}(R+a,T).
\label{eq17}
\end{equation}
\noindent
Using Eq.~(\ref{eq7}) this can be rearranged to the form
\begin{equation}
\frac{\partial F_{\rm sp}(a,T)}{\partial a}=
-2\pi RP(a,T)\left[1+\frac{J(a,T)}{RP(a,T)}\right],
\label{eq18}
\end{equation}
\noindent
where the following notation is introduced:
\begin{eqnarray}
&&
J(a,T)\equiv\frac{k_B T}{2\pi}
\sum_{l=0}^{\infty}{\vphantom{\sum}}^{\prime}
\sum_{n=1}^{\infty}\frac{1}{n}
\int_{0}^{\infty}k_{\bot}dk_{\bot}
\sum_{\alpha}r_{\alpha}^{2n}
\nonumber \\
&&~~~~~~~\times
\left[e^{-2q_lna}-
e^{-2q_ln(R+a)}\right].
\label{eq19}
\end{eqnarray}

Similar to Sec.~II, it can  easily be shown that at large $R$,
with $a$ and $T$ fixed,
it holds $J(a,T)/RP(a,T)\sim C/R$.
On the other hand, in the nonrelativistic limit
\begin{equation}
P(a,T)=-\frac{H}{6\pi a^3},
\quad
J(a,T)=\frac{RH(R+2a)}{12\pi a^2(R+a)^2},
\label{eq20}
\end{equation}
\noindent
where $H$ is defined in Eq.~(\ref{eq11}). {}From this it follows that
$J(a,T)/RP(a,T)\sim Ca$ when $a$ vanishes and $R$ is kept constant.
Thus, Eq.~(\ref{eq18}) can be rewritten in an equivalent form
\begin{equation}
\frac{\partial F_{\rm sp}(a,T)}{\partial a}=
-2\pi RP(a,T)\left[1+p(a,T)\frac{a}{R}\right],
\label{eq21}
\end{equation}
\noindent
where $p(a,T)$ is scarcely affected by $R$. Below we demonstrate that
in wide ranges of experimental parameters $p(a,T)$ is a very slowly
varying function and $|p(a,T)|<1/2$. Because of this, under the experimental
condition $a\ll R$ we can neglect the term of order $a/R$ in Eq.~(\ref{eq21})
and arrive at the equality
\begin{equation}
\frac{\partial F_{\rm sp}(a,T)}{\partial a}=
-2\pi RP(a,T),
\label{eq22}
\end{equation}
\noindent
which is the simplified formulation of the PFA for a gradient of the
Casimir force in sphere-plate configuration.
Equation (\ref{eq22}) was used for the comparison of experiment with
theory in dynamic measurements of the Casimir pressure
\cite{17,18,19,20,21,66}.

We have performed numerical computations of the quantity $p(a,T)$ as
a function of sepation and temperature for a sphere of fixed
radius $R=100\,\mu$m. In Fig.~3(a) the computational results for
$p(a,T)$ as a function of $a$ are presented at $T=300\,$K. Figure~3(b)
shows $p(a,T)$ as a function of $T$ at $a=2\,\mu$m. In both cases dotted,
dashed and solid lines indicate the use of ideal metal surfaces and
Au surfaces described by the Drude and plasma model approaches.
As can be seen in Fig.~3(a), within the separation region
from 100\,nm to $5\,\mu$m the function $|p(a,T)|$ varies between 0.31
and 0.48. When the temperature increases from absolute zero to $T=300\,$K,
$|p(a,T)|$ remains sandwiched between 0.31 and 0.39 [see Fig.~3(b)].
This demonstrates that possible  corrections to the simplified
formulation of the PFA (\ref{eq22}) do not exceed $a/R$.

It can be easily seen that the magnitude of $p(a,T)$ almost does not
depend on the radius of the sphere.  In Fig.~4 we present the computational
results for $p(a,T)$ as a function of $\log_{10}(a/R)$ at $T=300\,$K.
The groups of lines numbered 1 and 2 are for the separations
$a=0.1$ and $5\,\mu$m, respectively. The dotted, dashed and solid lines
have the same meaning as in Figs.~1--3. As is seen in Fig.~4, for the
group of lines 1 there is no dependence on $a/R$ over the whole range
from $10^{-4}$ to 0.05. Thus, at $a=0.1\,\mu$m, $p(a,T)$ does not depend
on $R$ for $R\geq 2\,\mu$m. For the group of lines 2, $p(a,T)$ does not depend
on $R$ for $10^{-4}\leq a/R\leq 10^{-3}$. At $a=5\,\mu$m this leads to
$R\geq 5000\,\mu$m. {}From Fig.~4 it also follows that relative magnitudes
of the function $p(a,T)$ computed using different models of dielectric
properties of metal depend on separation.

The experimental constraint on the magnitude of the
correction to Eq.~(\ref{eq22})  was obtained
in Ref.~\cite{65} by using several spheres
with different radii (see Sec.~II).
In the separation region $a<300\,$nm at $T=300\,$K it was shown that
$|p(a,T)|<0.4$ at a 95\% confidence level. This is in very good agreement
with the computational results in Fig.~4 and provides
the experimental confirmation
for the simplified formulation of the PFA in application to the gradient
of the thermal Casimir force between a sphere and a plate.

\section{Relationship between the proximity force
approximation and exact results for ideal metals}

In what follows we apply the PFA to calculate the thermal Casimir force
between a sphere and a plate made of ideal metal and discuss the
possibility to describe the thermal correction using this approximate
method. Using the notation (\ref{eq8}), Eq.~(\ref{eq4}) for the
thermal Casimir force can be presented in the form
\begin{eqnarray}
&&
F_{\rm sp}(a,T)=2\pi R{\cal F}_{\rm pp}(a,T)+2\pi I(a,T)
\label{eq23} \\
&&~~~~
\equiv 2\pi R{\cal F}_{\rm pp}(a,T)+2\pi X(a,T)-2\pi X(R+a,T).
\nonumber
\end{eqnarray}
\noindent
For ideal metals, using the dimensionless variables
\begin{equation}
y=2aq_l,\qquad \frac{2a\xi_l}{c}\equiv\tau_al
\label{eq24}
\end{equation}
in Eq.~(\ref{eq8}), the quantity $X(a,T)$ is given by
\begin{eqnarray}
X(a,T)&=&\frac{k_BT}{4\pi a}
\sum_{l=0}^{\infty}{\vphantom{\sum}}^{\prime}\sum_{n=1}^{\infty}
\frac{1}{n^2}\int_{\tau_al}^{\infty}dy\,e^{-ny}
\nonumber \\
&=&
\frac{k_BT}{8\pi a}\sum_{l=-\infty}^{\infty}
\int_{\tau_a|l|}^{\infty}dy\,\mbox{Li}_2(e^{-y}),
\label{eq25}
\end{eqnarray}
\noindent
where ${\rm Li}_n(z)$ is a polylogarithm function.

Equation (\ref{eq25}) can be presented in an equivalent form
convenient for the transition to the low-temperature limit using the
Poisson summation formula \cite{5,6}. According to this formula,
if $c(\alpha)$ is the Fourier transform of a function $b(x)$, i.e.
\begin{equation}
c(\alpha)=\frac{1}{2\pi}\int_{-\infty}^{\infty}b(x)\,e^{-i\alpha x}dx,
\label{eq26}
\end{equation}
\noindent
then it follows that
\begin{equation}
\sum_{l=-\infty}^{\infty}b(l)=
2\pi\sum_{l=-\infty}^{\infty}c(2\pi l).
\label{eq27}
\end{equation}
\noindent

Now we return to Eq.~(\ref{eq25}) and put
\begin{equation}
b(l)=\frac{k_BT}{8\pi a}\int_{\tau_a|l|}^{\infty}dy\,{\rm Li}_2(e^{-y}).
\label{eq28}
\end{equation}
\noindent
Keeping in mind that $b(l)=b(-l)$ and using Eq.~(\ref{eq26}),
one obtains
\begin{eqnarray}
&&
c(2\pi l)=\frac{1}{\pi}\int_{0}^{\infty}b(x)\,\cos(2\pi lx)\,dx
\label{eq29} \\
&&~~~~
=\frac{\hbar c}{32\pi^3a^2}\int_{0}^{\infty}dv\,\cos(lt_av)
\int_{v}^{\infty}dy\,{\rm Li}_2(e^{-y}).
\nonumber
\end{eqnarray}
\noindent
Here, we have introduced a new variable $v=\tau_a x$ and the notation
$t_a\equiv T_a/T$, where $k_BT_a\equiv\hbar c/(2a)$ is the effective
temperature related to the separation distance between the sphere and
the plate. Note that in terms of this notation it holds
$\tau_a=2\pi/t_a=2\pi T/T_a$. Taking into account that the quantity
$c(2\pi l)$ is an even function of its argument and using
Eqs.~(\ref{eq27}) and (\ref{eq29}), we arrive at
\begin{eqnarray}
&&
X(a,T)=4\pi\sum_{l=0}^{\infty}{\vphantom{\sum}}^{\prime}
c(2\pi l)
\label{eq30} \\
&&~~~
=\frac{\hbar c}{8\pi^2a^2}
\sum_{l=0}^{\infty}{\vphantom{\sum}}^{\prime}
\int_{0}^{\infty}dy\,{\rm Li}_2(e^{-y})
\int_{0}^{y}dv\,\cos(lt_av),
\nonumber
\end{eqnarray}
\noindent
where we exchanged the order of integrations with respect to $v$
and to $y$.

Equation (\ref{eq30}) can be represented in the form
\begin{equation}
X(a,T)=\frac{\hbar c}{8\pi^2a^2}\left[
\frac{1}{2}X_0+\sum_{l=1}^{\infty}X_l\right],
\label{eq31}
\end{equation}
\noindent
where
\begin{eqnarray}
X_0&=&\int_{0}^{\infty}ydy\,{\rm Li}_2(e^{-y}),
\label{eq32} \\
X_l&=&\frac{1}{lt_a}
\int_{0}^{\infty}dy\,{\rm Li}_2(e^{-y})\,
\sin(lt_ay).
\nonumber
\end{eqnarray}
\noindent
Direct calculation leads to
\begin{eqnarray}
&&
X_0=\sum_{n=1}^{\infty}\frac{1}{n^4}
\int_{0}^{\infty}xdx\,e^{-x}=\frac{\pi^4}{90},
\label{eq33}\\
&&
X_l=\frac{1}{lt_a}
\sum_{n=1}^{\infty}\frac{1}{n^3}
\int_{0}^{\infty}dx\,e^{-x}
\sin\frac{lt_ax}{n}
\nonumber \\
&&~
=
\sum_{n=1}^{\infty}\frac{1}{n^2(n^2+l^2t_a^2)}
=
\frac{1}{2l^4t_a^4}+\frac{\pi^2}{6l^2t_a^2}-
\frac{\pi}{2}\,\frac{\coth(\pi lt_a)}{l^3t_a^3},
\nonumber
\end{eqnarray}
\noindent
where $x=ny$. Substituting Eq.~(\ref{eq33}) into Eq.~(\ref{eq31}),
we obtain
\begin{eqnarray}
&&
X(a,T)=\frac{\pi^2\hbar c}{1440a^2}\left[1+
\frac{5}{t_a^2}
\vphantom{\sum_{l=1}^{\infty}}\right.
\label{eq34} \\
&&~~~~~~~
\left.-\frac{90}{\pi^3t_a^3}
\sum_{l=1}^{\infty}\frac{\coth(\pi lt_a)}{l^3}+
\frac{1}{t_a^4}\right].
\nonumber
\end{eqnarray}
\noindent
In the low-temperature limit it holds $T\ll T_a$, $t_a\gg 1$
and
Eq.~(\ref{eq34}) results in
\begin{eqnarray}
&&
X(a,T)=\frac{\pi^2\hbar c}{1440a^2}\left[1+
5\left(\frac{T}{T_a}\right)^2\right.
\label{eq35} \\
&&~~~~~~~~~~~~~~~~\left.
-\frac{90\zeta(3)}{\pi^3}
\left(\frac{T}{T_a}\right)^3+\left(\frac{T}{T_a}\right)^4\right],
\nonumber
\end{eqnarray}
\noindent
where $\zeta(z)$ is the Riemann zeta function.

The quantity $X(R+a,T)$ in Eq.~(\ref{eq23}) is given by Eq.~(\ref{eq34})
where $a$ is replaced with $R+a$, $t_a$ is replaced with
$t_R=T_R/T$ and $T_R$ is defined from
\begin{equation}
k_BT_R=\frac{\hbar c}{2(R+a)}\approx \frac{\hbar c}{2R}.
\label{eq36}
\end{equation}
\noindent
The sphere radius introduces a second characteristic temperature into
the problem which is much less than $T_a$ in experimentally relevant
situations. As an example, for typical spheres used in experiments on
measuring the Casimir force $R=100\,\mu$m and $a=100\,$nm resulting in
$T_R=11.4\,$K and $T_a=11400\,$K. Thus, for experiments performed at
room temperature, $T=300\,$K, it holds $T\ll T_a$, but $T\gg T_R$, i.e.,
a high-temperature regime with respect to the sphere radius, but a
low-temperature regime with respect to
the separation between the sphere and
the plate.

Nevertheless, we begin with the case of extremely low temperature with
respect to both parameters, $T\ll T_R\ll T_a$, which is achieved only
well below 1\,K. In this case the quantity $X(R+a,T)$ is also given by
Eq.~(\ref{eq35}) where $a$ is replaced with $R+a$ and $T_a$ is replaced
with $T_R$. We also take into account that under the condition
$T\ll T_a$ the free energy between two plane parallel plates in
Eq.~(\ref{eq23}) is given by \cite{5,6}
\begin{equation}
{\cal F}_{\rm pp}(a,T)=-\frac{\pi^2\hbar c}{720a^3}\left[1+
\frac{45\zeta(3)}{\pi^3}
\left(\frac{T}{T_a}\right)^3-\left(\frac{T}{T_a}\right)^4\right].
\label{eq37}
\end{equation}
\noindent
Substituting Eq.~(\ref{eq37}) and Eq.~(\ref{eq35}) (with the characteristic
temperatures $T_a$ and $T_R$) into Eq.~(\ref{eq23}), we arrive at the
result
\begin{equation}
{F}_{\rm sp}(a,T)=-\frac{\pi^3\hbar cR}{360a^3}\left[1-
\frac{a}{2R}+
\frac{a^3}{2R^3}\left(\frac{T}{T_R}\right)^4\right].
\label{eq38}
\end{equation}
\noindent
Here, we have included the main $T$-independent term of order $a/R$
and the first nonvanishing term depending on temperature.
The latter is of order
$T^4$ and is multiplied by the third power of the small parameter $a/R$.
For typical experimental parameters mentioned above $a/R\sim 10^{-3}$.
If one puts $T=1\,$K [the highest temperature where Eq.~(\ref{eq38}) is
applicable] the $T$-dependent term on the right-hand side of Eq.~(\ref{eq38})
appears to be of order $10^{-13}$.

Equation~(\ref{eq38}) can be identically rewritten in the form
\begin{equation}
{F}_{\rm sp}(a,T)=-\frac{\pi^3\hbar cR}{360a^3}\left(1-
\frac{a}{2R}\right)
+\Delta_T{F}_{\rm sp}(a,T),
\label{eq39}
\end{equation}
\noindent
where the thermal correction to the Casimir force is given by
\begin{equation}
\Delta_T{F}_{\rm sp}(a,T)=-\frac{\pi^3R^2(k_BT)^4}{45(\hbar c)^3}.
\label{eq40}
\end{equation}

It is interesting to compare this result obtained using the general
formulation of the PFA with the exact result for the thermal correction
in sphere-plate configuration \cite{67}.
In the functional determinant representation the exact free energy of
the Casimir interaction in sphere-plane geometry can be written as
\begin{equation}
{\cal F}_{\rm sp}^{\rm exact}(a,T)=\frac{k_BT}{2}
\sum_{l=-\infty}^{\infty}{\rm Tr}\,\ln\left[1-M(a,\xi_l)\right],
\label{eq41}
\end{equation}
\noindent
where the explicit form of the matrix $M$ in ideal metal case is
contained in Ref.~\cite{67}. In the limit of extremely low temperature,
$T\ll T_R$, and small separation distances, $a\ll R$, Ref.~\cite{67}
arrives at the following exact result for the temperature-dependent
part of the free energy
\begin{equation}
\Delta_T{\cal F}_{\rm sp}^{\rm exact}(a,T)=
\frac{\pi^3R^3}{225}\frac{(k_BT)^4}{(\hbar c)^3}
\left(\frac{29}{3}+\frac{112}{5}\,\frac{a}{R}\right).
\label{eq42}
\end{equation}
\noindent
{}From this equation, the exact thermal correction to the Casimir force
in the limiting case under consideration is the following
\begin{equation}
\Delta_T{F}_{\rm sp}^{\rm exact}(a,T)=
-\frac{\partial\Delta_T{\cal F}_{sp}^{\rm exact}(a,T)}{\partial a}=
-\frac{112\pi^3R^2}{1125}\frac{(k_BT)^4}{(\hbar c)^3}.
\label{eq43}
\end{equation}
\noindent
It can be seen from the comparison of Eqs.~(\ref{eq40}) and (\ref{eq43})
that although the most general formulation of the PFA gives the correct
dependence of the thermal correction on the fourth power of $T$, the
numerical coefficient before it is underestimated by a factor of 4.48.
At the same time, the application of the simplified formulation of the
PFA (\ref{eq13}) in this limiting case results in
an incorrect dependence
of the thermal correction on $T$ (the third power of $T$ instead of the
fourth). These results are not surprising. As explained in Sec.~II,
the PFA is applicable only under the condition $a\ll R$ and leads to
reliable predictions which are of zeroth order in the small parameter
$a/R$. Because of this, any contribution either to the free energy or
to the force which relative magnitude is
numerically of about $a/R$ or smaller calculated
using the PFA must be disregarded as physically meaningless.
Keeping in mind that in the limiting case of extremely low temperature,
$T\ll T_R$, the relative magnitude of
the thermal correction is much less than
$a/R$, the thermal effect in the Casimir force in this case should be
considered as unobservable.

A completely different type of situation occurs under the condition
$T\gg T_R$ which is well satisfied in all experiments on measuring the
Casimir force performed at room temperature (we remind that for the
sphere of $R=100\,\mu$m radius $T_R=11.4\,$K). Here, irrespective of
whether the inequality $T\ll T_a$ or $T\gg T_a$ is valid, the exact
expression (\ref{eq41}) leads in the zeroth order of the small parameter
$a/R$ to the simplified formulation of the PFA for the thermal
correction  to the Casimir force between a sphere and a plate \cite{67}:
\begin{equation}
\Delta_T{F}_{\rm sp}^{\rm exact}(a,T)=
2\pi R\Delta_T{\cal F}_{\rm pp}(a,T).
\label{eq44}
\end{equation}
\noindent
In the case $T\ll T_a$ the thermal correction to the Casimir free energy for
two plane parallel plates,
$\Delta_T{\cal F}_{\rm pp}(a,T)$, is contained in Eq.~(\ref{eq37}):
\begin{equation}
\Delta_T{\cal F}_{\rm pp}(a,T)=
-\frac{\hbar c\zeta(3)}{16\pi a^3}\left(\frac{T}{T_a}\right)^3
\left(1-\frac{T}{T_a}\right).
\label{eq45}
\end{equation}
\noindent
Thus, under a condition that $T\gg T_R$ the simplified formulation of
the PFA is applicable not only to the total quantities
$F_{\rm sp}(a,T)$ and ${\cal F}_{\rm pp}(a,T)$ [see Eq.~(\ref{eq13})],
but separately  to the zero-temperature and thermal contributions to
these quantities as well. This invalidates the statement of Ref.~\cite{62}
that for a sphere above a plate
at $T>T_R$ the PFA is inapplicable to each contribution
alone made on the basis of the worldline approach.

\section{Applicability of the proximity force
approximation to describe thermal correction to the Casimir
force between real metals}

Now we consider the possibility to apply the simplified formulation of
the PFA for the theoretical description of the thermal correction to the
Casimir force in recent experiments. As noted in the Introduction,
the thermal effect in the Casimir force is fundamentally different
depending on what model of the dielectric permittivity of metal (Drude or
plasma) is used. It is most straightforward to illustrate  this
difference in the low-temperature regime, $T\ll T_a$, for two plane
parallel plates described by the simple plasma model [Eq.~(\ref{eq16}) with
no contribution of core electrons, i.e., with $g_j=0$] or the Drude model
(\ref{eq15}). In the case when the simple plasma model is used the
temperature-dependent part of the free energy is given by \cite{68}
\begin{eqnarray}
&&
\Delta_T{\cal F}_{pp}^{(p)}(a,T)=
-\frac{\hbar c}{8\pi a^3}\left\{
\vphantom{\left[\left(\frac{T}{T_a}\right)^3\right] }
\frac{\zeta(3)}{3}\left(\frac{T}{T_a}\right)^3-
\frac{\pi^2}{90}\left(\frac{T}{T_a}\right)^4\right.
\nonumber\\
&&~~~~~
+\frac{\delta_0}{a}\left[\zeta(3)\left(\frac{T}{T_a}\right)^3
-\frac{2\pi^3}{45}\left(\frac{T}{T_a}\right)^4\right]
\label{eq46}\\
&&~~~~~~~~~~~\left.
-\left(\frac{\delta_0}{a}\right)^2\zeta(5)
\left(\frac{T}{T_a}\right)^5\right\},
\nonumber
\end{eqnarray}
\noindent
where $\delta_0=2\pi c/\omega_p$ is the effective skin depth in the
frequency range of infrared optics. If, however, the Drude model
is used under the condition $T\ll T_a$ one arrives at \cite{14}
\begin{eqnarray}
&&
\Delta_T{\cal F}_{pp}^{(D)}(a,T)=\Delta_T{\cal F}_{pp}^{(p)}(a,T)
\label{eq47} \\
&&~~~~~~~~~~~~~
+\frac{k_BT\zeta(3)}{16\pi a^2}\left[1-4\frac{\delta_0}{a}+
12\left(\frac{\delta_0}{a}\right)^2\right].
\nonumber
\end{eqnarray}
\noindent
As is seen in Eq.~(\ref{eq47}), the additional term emerging in the case
of the Drude model is positive and linear in temperature. Thus, it
dominates at low temperatures leading to
an anomalously large thermal
correction. Note that the condition of low temperature with respect to
$T_a$ is satisfied in all modern experiments performed at room
temperature at separations below $1\,\mu$m. At the same time for all
these experiments the condition of high temperature with respect
to $T_R$, i.e., $T\gg T_R$, is satisfied. Because of this, it can be
said \cite{67} that modern experiments on measuring the Casimir force
and its gradient in a sphere-plate configuration belong to the region
of medium temperatures.

The measurement data of all experiments performed in the configuration
of a rather large sphere in close proximity to a plane plate
(i.e., under the condition $a\ll R$) were compared with the theory using
the PFA. In static experiments (e.g., in
Refs.~\cite{17,29,30,48,49,50,51,54,55,56,57,58})
Eq.~(\ref{eq13}) was used for this purpose.
In dynamic experiments \cite{17,18,19,20,21,65,66} the form of the
PFA in Eq.~(\ref{eq22}) was employed. As was explained in Secs.~II and
III, in both cases terms of order $a/R$, as compared with unity, were
disregarded. This raises the question if this neglection imposes some
constraints on the possibility to observe thermal effects in the
Casimir force. Below we analyze this question using both theoretical
approaches to the calculation of
thermal Casimir force discussed in Sec.~II.

We start with the Drude model approach using the tabulated optical data
\cite{63} for ${\rm Im}\,\varepsilon(\omega)$ extrapolated  to low
frequencies by means of Eq.~(\ref{eq15}). The thermal effect can be
characterized by the relative thermal correction to the Casimir force
\begin{equation}
\delta_{T}^{(D,1)}(a,T)=
\frac{F_{\rm sp}^{(D)}(a,T)-F_{\rm sp}^{(D)}(a,0)}{F_{\rm sp}^{(D)}(a,T)}
\label{eq48}
\end{equation}
\noindent
and the Casimir pressure
\begin{equation}
\delta_{T}^{(D,2)}(a,T)=
\frac{P^{(D)}(a,T)-P^{(D)}(a,0)}{P^{(D)}(a,T)}.
\label{eq49}
\end{equation}
\noindent
We compute the relative correction (\ref{eq48}) by Eqs.~(\ref{eq5}) and
(\ref{eq13}) over the separation region from 0.1 to $5\,\mu$m. The
relative correction (\ref{eq49}) is computed over the same separation
region by using Eqs.~(\ref{eq3}) and
(\ref{eq5}). We remind that the measurement data for the quantity
$P(a,T)$ in dynamic experiments are obtained by using Eq.~(\ref{eq22})
from the gradient of the Casimir force
$\partial F_{\rm sp}(a,T)/\partial a$ which in its turn was recalculated
from the measured frequency shift. Thus, in static experiments using
sphere-plate configuration, the PFA in the form of Eq.~(\ref{eq13}) is
part of the theory, whereas in dynamic experiments
the PFA in the form of Eq.~(\ref{eq22}) allows to convert the experimental
data for the force gradient into the data for the equivalent Casimir
pressure.

In Fig.~5 we present the computational results in percents for the
relative thermal corrections $\delta_T^{(D,1)}$ (the solid line 1) and
$\delta_T^{(D,2)}$ (the solid line 2) at $T=300\,$K as a function of
separation in the region (a) from 0.1 to $5\,\mu$m and
(b) from 0.1 to $1\,\mu$m. In the same figure, the short-dashed and
long-dashed lines show the quantity $a/R$ in percents taken with the
same sign as the thermal corrections as a function of separation for
$R=100$ and $150\,\mu$m, respectively. These lines demonstrate the size
of typical errors arising from the use of the PFA.
As can be seen in Fig.~5, the relative thermal corrections for both the
Casimir force and equivalent Casimir pressure are rather large.
In the region from 1 to $4\,\mu$m they take negative values and achieve
--35\% and --47\%, respectively. Within the separation region
from 0.1 to $1\,\mu$m the magnitudes of  thermal corrections to the
force and to the pressure increase from 1.5\% and 0.7\% to 23\% and 16\%,
respectively. In the same separation region the error due to the use of
the PFA varies from 0.1\% to 1\% for the sphere of $100\,\mu$m radius
and from 0.07\% to 0.7\% for the sphere of $150\,\mu$m radius.
Thus, the use of the PFA in the Drude model approach does not impose
any constraints on the possibility to measure large thermal effects
predicted in this approach.

The answer to the question whether or not the thermal corrections
to the Casimir force in a sphere-plate configuration can be measured
depends not only on the errors due to the application of the PFA, but
also on the size of  the total experimental error. Thus, in the static
experiment \cite{51} performed by means of an atomic force microscope
the total relative experimental error of the Casimir force $F_{\rm sp}$
determined at a 60\% confidence level at separations 100 and 200\,nm is
equal to 3.5\% and 22.4\%, respectively \cite{69}. These are larger than
the relative thermal correction $\delta_T^{(D,1)}$ in Fig.~5(b).
Because of this, in the experiment of Refs.~\cite{51,69} the predicted
thermal correction cannot be either confirmed or excluded.
A completely different type of situation occurs in dynamic experiments
of Refs.~\cite{17,18,19,20,21} performed by means of a micromachined
oscillator. Thus, in the most precise experiment of Refs.~\cite{20,21}
the total relative experimental error of the Casimir pressure $P$
determined even at a higher, 95\%, confidence level varies from 0.19\%
to 0.9\% and to 9\% when separation increases from 160 to 400 and
to 750\,nm, respectively. As can be seen in Fig.~5(b), the magnitude of
the relative thermal correction $\delta_T^{(D,2)}$ remains much larger
than the total experimental error over the entire range of experimental
separations. That is the reason why, when no evidence of the predicted
thermal corrections was found, Refs.~\cite{20,21} arrived at the
conclusion that the Drude model approach is experimentally excluded
at a 95\% confidence level (within a narrower separation region from
210 to 620\,nm this approach was  excluded
at a 99.9\% confidence level \cite{21}).

It has been speculated in the literature that the experimental exclusion
of the Drude model approach might be not warranted because of some
theoretical and experimental uncertainties. Specifically, it was
stressed \cite{70} that the Drude parameters $\omega_p$ and $\gamma$
used for the extrapolation of the optical data to low frequencies may
vary. Besides, Ref.~\cite{71} surmised that measurements of
absolute separations in Ref.~\cite{48} (which data were also used to
exclude the Drude model approach) contain an unaccounted systematic error
which could bring the data away from the theory. Here, we make a test of
both these opportunities with respect to the theory-experiment
comparison in Refs.~\cite{20,21}. In Fig.~6 the differences between
the theoretical Casimir pressures computed using the Drude model approach,
as explained above but with account of surface roughness,
and the mean experimental Casimir pressures measured
in Refs.~\cite{20,21} are indicated as dots. The theoretical pressures
for the lower set of dots are computed with the conventional Drude
parameters of Sec.~II ($\omega_p=9.0\,$eV, $\gamma=0.035\,$eV).
The theoretical Casimir pressures
for the upper set of dots are computed with the alternative  Drude
parameters \cite{70}, $\omega_p=6.82\,$eV, $\gamma=0.0405\,$eV,
where the value of the plasma frequency is most different from the
conventional one. The differences $P^{\rm theor}-\bar{P}^{\rm expt}$ with
$P^{\rm theor}$ computed using all other alternative Drude parameters
listed in Ref.~\cite{70} are sandwiched between the two sets of dots
shown in Fig.~6.
The solid lines in Fig.~6 indicate the borders of the confidence
intervals $[-\Xi(a),\Xi(a)]$ computed in Ref.~\cite{21} at a 95\%
confidence level. By the construction of  $[-\Xi(a),\Xi(a)]$,
95\% of dots must belong to these intervals. However, as is seen in Fig.~6,
almost all dots in both sets lie outside the solid lines demonstrating
that the theoretical approach using the Drude model is excluded by the
data at a 95\% confidence level over the entire measurement range
(as was mentioned above, within a bit narrower reparation region the
exclusion at a 99.9\% confidence level holds). It is pertinent to note
also that the use of optical data of Ref.~\cite{70} for the first
absorption bands (which are slightly different from the data of
Ref.~\cite{63} used in our computations) does not change this conclusion.

Now we consider the possibility to bring the experimental data in
agreement with the Drude model approach by assuming that there is an
unaccounted systematic error in the determination of absolute separations.
Note that this systematic error (if any) is separation-independent.
This is because in the setup used the differences between the
values of separations where the Casimir pressures were measured
are fixed interferometrically to high precision (see Ref.~\cite{18}
for details). To find absolute separations, one should know the
initial absolute separation which is determined from the electrostatic
calibrations. If one admits that electrostatic calibrations contain
some uncertainty, the initial separation would be  burdened  with some
unaccounted constant systematic error which is translated
to all separations.
In an attempt to place the dots within the limits of the confidence
intervals, in Fig.~7(a) we plot the differences
$P^{\rm theor}-\bar{P}^{\rm expt}$ with two sets of the Drude parameters
mentioned above as a function of separation, but with all separation
distances decreased by $\Delta a=1\,$nm.
(Note that the increase of separations results in even larger
deviations of dots from the confidence intervals than in Fig.~6.)
This corresponds to the assumption that an unaccounted systematic error
in the measurement of separation distances in Refs.~\cite{20,21} is
equal to 1\,nm (remind that in actual fact the total error in the
measurement of separations in Refs.~\cite{20,21} is equal to 0.6\,nm).
As can be seen in Fig.~7(a), the shift of separations for 1\,nm does not
furnish the desired result for the lower set of dots at separations
above 230\,nm and is not helpful for the upper set of dots at any
separation. The decrease of all separations for 3\,nm [see Fig.~7(b)]
expels the lower set from the confidence intervals
at separations below 230\,nm, but not yet includes these dots into
the confidence interval at $a>310\,$nm. In Fig.~7(c) we illustrate the
largest decrease of separations for 6\,nm. Here, all dots related to the
lower set are still outside the confidence intervals at $a>430\,$nm,
but the dots of both sets are already outside of them at shortest
separations.
Note that similar to Fig.~6 the differences
$P^{\rm theor}-\bar{P}^{\rm expt}$ computed with all other
alternative Drude parameters are sandwiched between the two sets of
dots shown in each of Figs.~7(a), 7(b), and 7(c). The same is
correct when the optical data (not just the Drude parameters)
from different sources are used. As shown in Ref.~\cite{74a},
the use of some alternative optical data instead of the data of
Ref.~\cite{63} would decrease the magnitudes of theoretical
Casimir pressures and, thus, only increase discrepances between
the predicted and experimental Casimir pressures.
This means that no systematic error in the measurement
of absolute separations can bring the experimental data of
Refs.~\cite{20,21} in agreement with the Drude model approach using any
values of Drude parameters and any sets of the optical data.

{}From the above it follows that the sensitivity of dynamic experiments
by means of micromachined oscillator is quite sufficient to registrate
the thermal correction to the Casimir pressure, as predicted by the
Drude model approach. What is more, the application of the PFA in the
form of Eq.~(\ref{eq22}) in this case is warranted because both the
total quantities $\partial F_{\rm sp}/\partial a$ and $P$ and contained in
them thermal corrections are much larger than $a/R$ for the experimental
parameters. The lack of any observation effect for the thermal correction
to the Casimir pressure, as predicted by the Drude model approach, means
that this approach is experimentally inconsistent.

We are coming now to the question is it possible to observe small
thermal effect in the Casimir force, as predicted by the plasma model
approach in sphere-plate configuration. In this case the relative thermal
correction to the force, $\delta_T^{(p,1)}$, and to the pressure,
$\delta_T^{(p,2)}$, are expressed once again by Eqs.~(\ref{eq48}) and
(\ref{eq49}) where $F_{\rm sp}^{(D)}$ and $P^{(D)}$ are replaced with
$F_{\rm sp}^{(p)}$ and $P^{(p)}$ computed using the dielectric permittivity
of the generalized plasma-like model (\ref{eq16}). In Fig.~8 the
computational results (in percents) are presented for $\delta_T^{(p,1)}$
(the solid line 1) and for $\delta_T^{(p,2)}$
(the solid line 2) at $T=300\,$K as a function of $a$ in the region
(a) from 0.1 to $5\,\mu$m and (b) from 0.1 to $1\,\mu$m.
The short-dashed and long-dashed lines show the quantity $a/R$
(in percents) for $R=100$ and $150\,\mu$m, respectively.
As can be seen in Fig.~8, the thermal corrections computed using the
plasma model approach are positive and increase monotonously with the
increase of separation. At $a<1\,\mu$m their magnitudes are much less
than the magnitudes of the thermal corrections computed using the
Drude model approach  (see Fig.~5). As a result, $\delta_T^{(p,1)}$
lies below the short-dashed line related to the static experiments.
The relative correction to the Casimir pressure, $\delta_T^{(p,2)}$,
lies below the long-dashed line related to the dynamic experiments
at $a<1.6\,\mu$m. This means that one cannot attribute physical
meaning to such small values of the thermal corrections $\delta_T^{(p,1)}$
and $\delta_T^{(p,2)}$ computed using the PFA [Eqs.~(\ref{eq13}) and
(\ref{eq22})] which disregards contributions to the force and to the
pressure of order $a/R$.

The magnitudes of the thermal corrections $\delta_T^{(p,1)}$
and $\delta_T^{(p,2)}$ are markedly larger than $a/R$
shown by the short-dashed
(long-dashed) lines only at separations $a>0.6\,\mu$m ($a>2\,\mu$m).
However, either in the static experiment \cite{51} on measuring the
Casimir force $F_{\rm sp}$ or in the most precise dynamic experiment
\cite{20,21} on measuring the Casimir
pressure $P$, over the entire separations
regions $a\geq 0.1\,\mu$m and $a\geq 0.16\,\mu$m,
the magnitudes of $\delta_T^{(p,1)}$ and $\delta_T^{(p,2)}$
fall far short of the total relative experimental error.
It is not surprising, then, that thermal effects were not observed in
these experiments.  One more possibility for future measurements is to
use  spheres of larger radius. This allows to make $a/R$ smaller and,
thus, make thermal corrections $\delta_T^{(p,1)}$ and $\delta_T^{(p,2)}$
computed using the PFA physically meaningful over the entire range of
separations  $a\geq 0.1\,\mu$m. In addition, in static experiments the
Casimir force $F_{\rm sp}$ is larger for a sphere of larger $R$ resulting in
a smaller  relative experimental error. For spheres of centimeter-size
radii, however, inavoidable deviations from perfect spherical shape is
a problem of great concern which prevents accurate electrostatic
calibration \cite{72}. It is therefore unlikely that small thermal
effect in the Casimir force, as predicted by the plasma model approach,
will be observed in the configuration of a sphere above a plate.

We conclude this section with a brief discussion of recent exact results
for the thermal Casimir force between a sphere and a plate described by
simple plasma and Drude models without account for interband transitions
of core electrons \cite{23} and by the generalized plasma-like and
Drude-like
models \cite{24}. The key question is whether or not these results
support computations performed using the PFA. As was discussed in
Sec.~IV, for ideal metals  the exact results in the zeroth order in
$a/R$ coincide with the PFA if $T\gg T_R$. This is the case for both
the total Casimir force or Casimir pressure and separately for the
thermal corrections.
Unfortunately, the exact computations of Refs.~\cite{23,24} were
performed in regions far away from the values of experimental
parameters and outside the region where the PFA is applicable.
For example, in
Ref.~\cite{23} the sphere radii were chosen to be
$R=0.1,\> 0.2,\> 0.5,\> 1,\> 2,\> 5\,\mu$m and computations were performed
at separations from 0.5 to $10\,\mu$m. For a sphere radius $R=10\,\mu$m
the computations of Ref.~\cite{23} were made from $a=1$ to $10\,\mu$m.
Thus, in all cases considered it holds
$a/R\geq 0.1$. Remind that for the
experimental parameters of Refs.~\cite{51,69} $a/R$ varies from 0.00063
to 0.003 and for the  parameters of Refs.~\cite{20,21} from 0.0011 to 0.005.

According to the results presented in Secs.~II and III, the PFA in the
form of Eqs.~(\ref{eq13}) and (\ref{eq22}) is applicable only at
$a/R\ll 1$. Because of this, the speculation of Ref.~\cite{23} that at small
separations the Drude and plasma models lead to Casimir force values much
closer than predicted by the PFA made on the basis of computations
in the region of $a/R>0.1$ is unjustified. The computational results of
Ref.~\cite{23} at $a=0.5\,\mu$m, $T=300\,$K clearly demonstrate that
deviations between the exact theory and the PFA vanish with decreasing $a/R$.
Thus, from Fig.~3 of Ref.~\cite{23} it can be seen that relative
deviation of the quantity $F_{\rm sp}^{(p)}(a,T)/F_{\rm sp}^{(D)}(a,T)$
computed exactly from that computed using the PFA decreases from 9.2\%
to 2.5\% when the sphere radius increases from $R=0.1\,\mu$m ($a/R=5$)
to $R=5\,\mu$m ($a/R=0.1$). In such a manner exact computations confirm
that the computational error in the ratio
$F_{\rm sp}^{(p)}(a,T)/F_{\rm sp}^{(D)}(a,T)$
arising from using the PFA for real metals (0.025 for $a/R=0.1$)
is smaller that $a/R$. For the experimental values of $a/R$ mentioned
above this error would be much less than the total experimental error.
Because of this it can be stated with certainty that the
comparison of the performed experiments with the exact theory
instead of the PFA
(when the exact will become available) will not lead to
any changes with respect to already obtained conclusions.

Both papers \cite{23,24} claim that for large separations in the
sphere-plane geometry the Drude model leads to a force a factor of 3/2
smaller than the plasma model (instead of a factor of 2 as
it holds for two parallel metal plates).
In such a general form this formulation is, however,
somewhat misleading as it does not cover all relevant limiting cases.
In fact it is correct only in the case when
$\lambda_p=2\pi c/\omega_p\ll R\ll a$ \cite{23,24}, i.e., outside the
application region of the PFA. If, however, we consider large separations
$a\gg \hbar c/(2k_BT)$, i.e., $T\gg T_a$, and large sphere $R\gg a$
 (which
means that the PFA is applicable), then the exactly calculated ratio
$F_{\rm sp}^{(p)}/F_{\rm sp}^{(D)}=2$. The same result is obtained using
the PFA. As shown in Ref.~\cite{24} by means of exact computations,
there is one more limiting case, $a\gg R$, $R\leq \lambda_p$ and
$T\gg T_a$, where $F_{\rm sp}^{(p)}/F_{\rm sp}^{(D)}=1$. This result is
quite natural because a small sphere above a plate can be modelled
as an atom characterized by some effective dynamic polarizability \cite{73}.
Keeping in mind that at large separations only the static polarizability is
relevant which is common for the plasma and Drude models, both the free
energies and forces computed using these models coincide in the limit
of large $a$.
Thus, in sphere-plate geometry the ratio $F_{\rm sp}^{(p)}/F_{\rm sp}^{(D)}$
can take different values, 3/2, 2, and 1, in the large distance limit
depending on the parameters of the problem. In the experimental
situations, however, where $a\ll R$ and $R\gg\lambda_p$, in the limit
of large distances, $a\gg\hbar c/(2k_BT)$, it holds
$F_{\rm sp}^{(p)}/F_{\rm sp}^{(D)}=2$ like for two parallel plates.

In Ref.~\cite{24} the exact computations of the free energies
${\cal F}_{\rm sp}^{(p)}$ and ${\cal F}_{\rm sp}^{(D)}$ were performed
for a sphere of $R=5\,\mu$m radius at $T=300\,$K and 77\,K within the
separation region from 0.265 to $95\,\mu$m. This corresponds to the
values of $a/R$ varying from 0.053 to 19. Once again, only near the
smallest $a/R$ considered the PFA becomes applicable, whereas for the
comparison with recent experiments more than one order of magnitude
smaller ratios $a/R$ are required. At the shortest separation
$a=0.265\,\mu$m ($a/R=5.3$\%) from Fig.~5 of Ref.~\cite{24} one
obtains the exact value for the ratio
${\cal F}_{\rm sp}^{(p)}/{\cal F}_{\rm sp}^{(D)}=1.11$ at $T=300\,$K.
The same ratio computed using the PFA is equal to 1.14 leading to a
relative 2.6\% deviation between the PFA and exact results which is less
than one half of $a/R$. This confirms that for real metals
in sphere-plate geometry the
PFA results for both the thermal Casimir force and
 free energy  go to the respective exact results with
decreasing $a/R$.

\section{The possibility to measure thermal effect in the
{\protect \\} gradient of the Casimir pressure}

The original Casimir configuration of two parallel plates is the only
one where  the comparison of experiment with theory does not require
either the PFA or more sophisticated exact computational methods which
can be effectively used so far solely in some restricted ranges of
parameters. The only experiment of this kind was performed \cite{74} in the
modern stage of the Casimir force measurements. This is an experiment
of dynamic type where one of the plates was oscillating at the natural
oscillator frequency. As a result, the shift of this frequency
under the influence of the Casimir pressure was measured and recalculated
into the gradient of the Casimir pressure. The experiment \cite{74}
reported 15\% measure of agreement between the data and theory using
the model of ideal metals (see review \cite{11} for details).

The gradient of the Casimir pressure
\begin{equation}
P^{\prime}(a,T)\equiv\frac{\partial P(a,T)}{\partial a}
\label{eq50a}
\end{equation}
\noindent
is obtained from Eqs.~(\ref{eq3})
and (\ref{eq5}) by differentiation with respect to separation
\begin{equation}
P^{\prime}(a,T)
=\frac{2k_BT}{\pi}\sum_{l=0}^{\infty}
{\vphantom{\sum}}^{\prime}\int_{0}^{\infty}\!\!q_l^2k_{\bot}
dk_{\bot}\sum_{\alpha}
\frac{r_{\alpha}^2e^{-2q_la}}{(1-r_{\alpha}^2e^{-2q_la})^2}.
\label{eq50}
\end{equation}
\noindent
The thermal correction to the gradient of the Casimir pressure
can be characterized by the quantity
\begin{equation}
\delta_{T}^{(3)}(a,T)=
\frac{P^{\prime}(a,T)-P^{\prime}(a,0)}{P^{\prime}(a,T)}.
\label{eq51}
\end{equation}
\noindent
We will add additional indices $D$, $p$ and $IM$ and calculate the
quantities $\delta_{T}^{(D,3)}$, $\delta_{T}^{(p,3)}$ and
$\delta_{T}^{(IM,3)}$ depending on the used model of dielectric
properties of metal described in Sec.~II.

The computational results for the relative thermal correction (\ref{eq51})
to the gradient of the Casimir pressure as a function of separation
at $T=300\,$K are presented in Fig.~9. Here, the separation region from
0.5 to $5\,\mu$m is considered because at shorter separations it would be
hard to experimentally keep the plates parallel.
The dashed, solid, and dotted lines represent results computed by using
the tabulated optical data extrapolated to low frequencies by the Drude
model with conventional parameters, the generalized plasma-like model,
and the model of ideal metals, respectively. As can be seen in Fig.~9,
for ideal metal plates and plates described by the generalized plasma-like
model the relative thermal correction to the gradient of the Casimir
pressure monotonously increases with the increase of separation.
At separations $a\leq 3\,\mu$m, where the experiment is feasible,
$\delta_T^{(p,3)}\leq 3.6$\% and $\delta_T^{(IM,3)}\leq 1.3$\%.
So small magnitudes of $\delta_T^{(p,3)}$ make it impossible to observe the
thermal corrections due to the generalized plasma-like model in the
measurements of the gradient of the Casimir pressure.
By contrast, the relative thermal correction $\delta_T^{(D,3)}$
computed using the Drude model approach decreases monotonously with
increase of separation from 0.5 to $4.1\,\mu$m where it achieves the
minimum value of --53.3\%. At $a=3\,\mu$m it holds
$\delta_T^{(D,3)}=-43$\%. Thus, the proposed experiment on measuring the
gradient of the Casimir pressure \cite{75} might provide one more
confirmation for the exclusion of the Drude model approach, but is
not well adapted for the measurement of the thermal correction, as
predicted by the generalized plasma-like model.
Note that according to the authors of Ref.~\cite{75} their effort
to measure the Casimir force is not yet successful due to the use
of Al surfaces. The point is that Al is the subject of quick
oxidation even in high vacuum. For this reason, in the first
precise measurement of the Casimir force \cite{49}, Al surfaces
were coated with thin transparent layers of Au/Pd. The use of such
layers, however, complicates the comparison between experiment and
theory. As a result, starting from 2000 almost all experiments
on the Casimir force between metallic test bodies \cite{11}
used Au coated surfaces. Presently another setup with parallel
Au surfaces, instead of Al, is under construction \cite{75}.

It can be shown also analytically that measurements of the gradient of
the Casimir pressure are unsuitable for the detection of small thermal
corrections. For example, for ideal metal plates in the limit of
low temperatures $T\ll T_a$ it holds \cite{6}
\begin{equation}
P^{(IM)}(a,T)=-\frac{\pi^2\hbar c}{240 a^4}\left[1+\frac{1}{3}
\left(\frac{T}{T_a}\right)^4-\frac{120}{\pi}\frac{T}{T_a}
e^{-2\pi T_a/T}\right].
\label{eq52}
\end{equation}
\noindent
Taking into account that the second contribution on the right-hand side
of Eq.~(\ref{eq52}) does not depend on separation, the main contribution
to the gradient of the Casimir pressure in the low-temperature limit takes
the form
\begin{equation}
\frac{\partial P^{(IM)}(a,T)}{\partial a}=
\frac{\pi^2\hbar c}{60 a^5}\left(1+
60e^{-2\pi T_a/T}\right).
\label{eq53}
\end{equation}
\noindent
As can be seen on the right-hand side of Eq.~(\ref{eq53}),
there are no terms in powers of $T/T_a$ and the thermal correction is
exponentially small. Equation (\ref{eq53}) reproduces the exactly calculated
$\partial P^{(IM)}(a,T)/\partial a$ at $T=300\,$K with an error of less than
1\% up to the separation distance $a=3\,\mu$m. The absence of power-type
thermal corrections in the gradient of the Casimir pressure for ideal
metals explains why the thermal effect in this case is so small.

At low temperatures, the analytical representation for the gradient of
the Casimir pressure can be obtained for real metals as well.
Thus, for metals described by the generalized plasma-like model
at separations above $0.5\,\mu$m considered here the influence of
interband transitions can be neglected.
Then the thermal correction in the gradient of the Casimir pressure is
obtained as the negative second derivative of
Eq.~(\ref{eq46}) leading to
\begin{eqnarray}
&&
\frac{\partial P^{(p)}(a,T)}{\partial a}=
\frac{\pi^2\hbar c}{60 a^5}\left[1-
\frac{20}{3}\frac{\delta_0}{a}+ 36\left(\frac{\delta_0}{a}\right)^2
\right.
\nonumber \\
&&~~~~~~~~\left.
+
\frac{15\zeta(3)}{\pi^3}\frac{\delta_0}{a}
\left(\frac{T}{T_a}\right)^3\right],
\label{eq54}
\end{eqnarray}
\noindent
where the exponentially small terms are omitted.
Here, the leading thermal correction is of power-type, but it contains
the first and third power of small parameters.
 Equation (\ref{eq54}) provides a better than 1\% approximation for
the exactly computed $\partial P^{(p)}(a,T)/\partial a$ over the range of
separations below $1.3\,\mu$m.
Similar low-temperature results for metals described by the Drude model
can be obtained by calculating the negative second derivative of
Eq.~(\ref{eq47})
\begin{equation}
\frac{\partial P^{(D)}(a,T)}{\partial a}=
\frac{\partial P^{(p)}(a,T)}{\partial a}-
\frac{3k_BT\zeta(3)}{8\pi a^4}\left(1-
8\frac{\delta_0}{a}+ 40\frac{\delta_0^2}{a^2}\right)
\label{eq55}
\end{equation}
\noindent
This approximate expression reproduces the values of
$\partial P^{(D)}(a,T)/\partial a$ with an error less than 1\% at
separations below $1\,\mu$m.

To conclude this section, for the configuration of two parallel plates
in the dynamic regime the relative thermal correction to the gradient
of the Casimir pressure computed using the plasma model approach
achieves 20\% only at $a=5\,\mu$m (see Fig.~9). The same relative size of the
thermal correction computed using the plasma model but to the Casimir
pressure is achieved at a shorter separation $a=3.7\,\mu$m
[see the solid line 2 Fig.~8(a)]. Thus, the static experiment with two
parallel plates of sufficiently large area might be preferable for
the registration of small thermal corrections at relatively large
separations.  Although at the same separation the relative thermal
correction to the Casimir force in a sphere-plate configuration is
larger [43\% in accordance with  the solid line 1 in Fig.~8(a)
at $a=3.7\,\mu$m], small values of forces for spheres of relatively
small radii make its measurement impossible. As was discussed above, the
use of centimeter-size spheres meets difficulties due to unavoidable
deviations from sphericity. In this respect the increase of plate area
and, thus, the increase of the Casimir force, may make possible to
decrease the experimental error at separations of a few micrometers to
about 10\%, i.e., below the characteristic size of small thermal
effect predicted by the plasma model approach.

\section{Conclusions and discussion}

In the foregoing we have analyzed the possibility to measure
thermal effects in the Casimir force between metals in the
configurations of a sphere above a plate and two parallel plates.
In connection with the sphere-plate configuration, a detailed study
of the reliability of the PFA to describe the thermal Casimir force
was performed. We have considered the two formulations of the PFA,
the most general, Derjaguin, formulation \cite{46} and the simplified
formulation of Ref.~\cite{22} which was used for the comparison of many
experiments on the measuring of the Casimir force with theory.
We show in Sec.~II that for a large sphere, i.e., under the condition
$a\ll R$, the magnitude of the relative difference between thermal
Casimir forces $F_{\rm sp}$ computed for real metals by using the two
formulations of the PFA is less than $a/R$. This holds regardless of what
approach to the description of charge carriers (Drude or plasma model)
is employed. A similar result is obtained in Sec.~III for the gradient
of the thermal Casimir force in sphere-plate configuration,
$\partial F_{\rm sp}/\partial a$. Here again the relative difference
computed using the two formulations is less than $a/R$.
The obtained results are compared with the experimental data of
Ref.~\cite{65} demonstrating that for large spheres of different radii
at $T=300\,$K the relative differences of the measured
$|F_{\rm sp}|$, $\partial F_{\rm sp}/\partial a$ and those computed
using the simplified formulation of the PFA are also less than $a/R$.
{}From this we can conclude that the simplified formulation of the PFA
is a good approximation for the calculation of thermal Casimir forces
in sphere-plate configuration for spheres of large radii.

In Sec.~IV special attention is paid to the case of an ideal metal sphere
above an ideal metal plate. In this case a simple analytic expression for
the difference between the two formulations of the PFA is obtained.
The thermal correction to the Casimir force computed using the PFA
is considered both in the limiting case of very low temperatures,
$T\ll T_R$, and in the case $T\gg T_R$. The latter includes the region
of parameters where all measurements of the Casimir force in
sphere-plate configuration have been performed. The results for the
thermal correction obtained using the PFA are compared with available
exact results. It is emphasized that in the experimental region
$T\gg T_R$ the exact results of Ref.~\cite{67} in the zeroth order of
$a/R$ coincide with the simplified formulation of the PFA.
This provides a theoretical basis for the application of the PFA in the
measurements of the thermal Casimir force for the comparison between
experiment and theory.

The applicability of the PFA and the possibility to observe the thermal
correction to the Casimir force in the configuration of a sphere above
a plate made of real metals are discussed in Sec.~V.
Here, the relative thermal corrections to the Casimir force and to
the Casimir pressure are computed in the framework of both the Drude
model and the plasma model approaches. The obtained results are
compared with the error introduced from the use of the PFA and with
typical experimental errors of force and pressure measurements.
New evidence is provided that some unaccounted systematic errors
in the experiment of Refs.~\cite{20,21} cannot bring the measurement
results in agreement with the prediction of the Drude model approach.
It is argued that the configuration of a sphere above a plate is well
suited to exclude the large thermal correction derived within the Drude
model approach but scarcely can be used for the experimental
observation of small thermal effects, as predicted from the plasma model
approach. The reason is that the respective thermal correction
$\delta_T^{(p)}$ is several times larger than the error of PFA,
$a/R$, only at sufficiently large separations, where the error of force
or equivalent pressure measurements exceeds $\delta_T^{(p)}$.
The exact computational results for the thermal correction to the
Casimir force between real metals described by the Drude and plasma
models are compared with the PFA results. It is shown that the region
of parameters where the exact results are available is outside the area
of application of the PFA. We demonstrate that at relatively large
separations the ratio $F_{\rm sp}^{(p)}/F_{\rm sp}^{(D)}$ can go
to different limiting values (2, 3/2 or 1) depending on the
relationship between the parameters $a$, $R$, $\lambda_p$ and $T$.
This generalizes the results of Refs.~\cite{23,24}. The performed
comparison enables us to conclude that the available exact results for
real metals converge to the PFA results when $a/R\to 0$.
For the experimental values of $a/R$ varying from 0.00063 to 0.005
the errors due to the use of the PFA instead of the exact methods
are much
less than the experimental error of force and pressure measurements
in all experiments performed to the present day.

In the configuration of two parallel plates, we have calculated the
relative thermal correction to the gradient of the Casimir pressure
which is measured in the dynamic regime (Sec.~VI). We show that a small
thermal effect, as predicted in the plasma model approach, is
additionally suppressed in the gradient of the Casimir pressure.
Because of this, the original Casimir configuration of two parallel
plates of sufficiently large area used in the static regime remains the
most prospective for measurement of small thermal effect in the
Casimir force.

\section*{Acknowledgments}
The authors acknowledge helpful discussions with M.~Bordag.
 G.L.K. and V.M.M. are
grateful to the  Institute
for Theoretical Physics, Leipzig University,
where this work was performed, for kind
hospitality.
They  were supported by Deutsche Forschungsgemeinschaft,
Grant No.~GE\,696/10--1.
G.L.K.\ was also supported by the Grant of the Russian Ministry of
Education P--184.


\begin{figure*}[h]
\vspace*{-4.cm}
\centerline{
\includegraphics{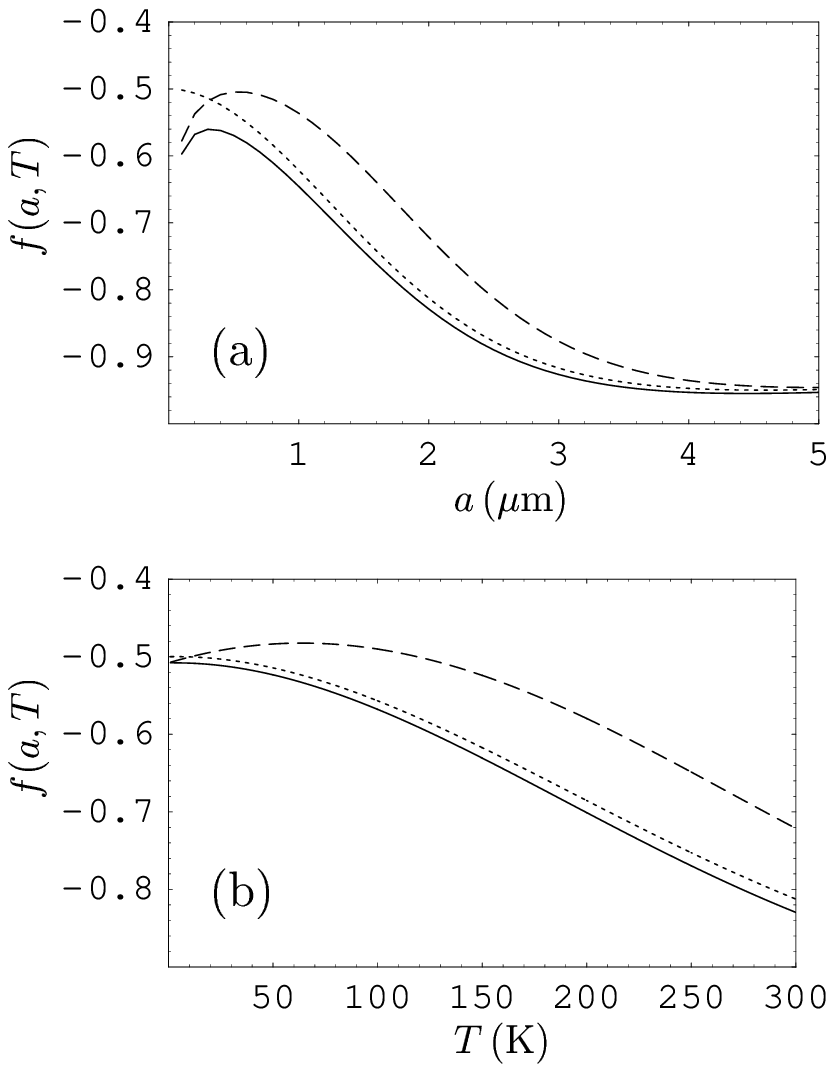}
} \vspace*{-11.cm} \caption{The quantity $f$ characterizing differences
between the two formulations of the PFA
for the Casimir force as a function of (a) separation
at room temperature $T=300\,$K and (b) temperature at a separation
$a=2\,\mu$m. The dotted, dashed and solid lines show the results
computed using ideal metals, Drude and plasma model approaches,
respectively. The sphere radius $R=100\,\mu$m.}
\end{figure*}
\begin{figure*}[h]
\vspace*{-8.cm}
\centerline{\hspace*{2cm}
\includegraphics{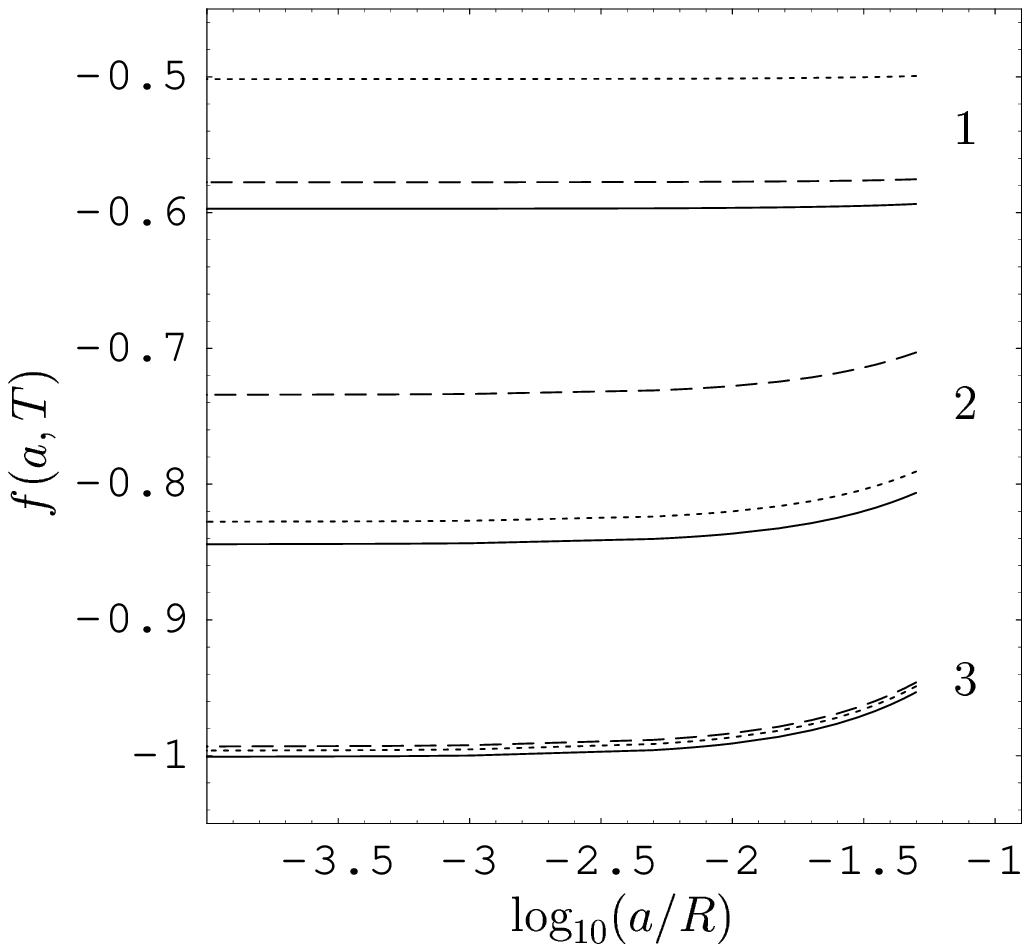}
} \vspace*{-6.5cm} \caption{The quantity $f$ characterizing differences
between the two formulations of the PFA
for the Casimir force as a function of $a/R$ at
room temperature $T=300\,$K.
The dotted, dashed and solid lines show the results
computed using ideal metals, Drude and plasma model approaches,
respectively.
The groups of lines numerated 1, 2, and 3 are for the respective
fixed separations $a=0.1$, 2, and $5\,\mu$m.}
\end{figure*}
\begin{figure*}[h]
\vspace*{-4.cm}
\centerline{
\includegraphics{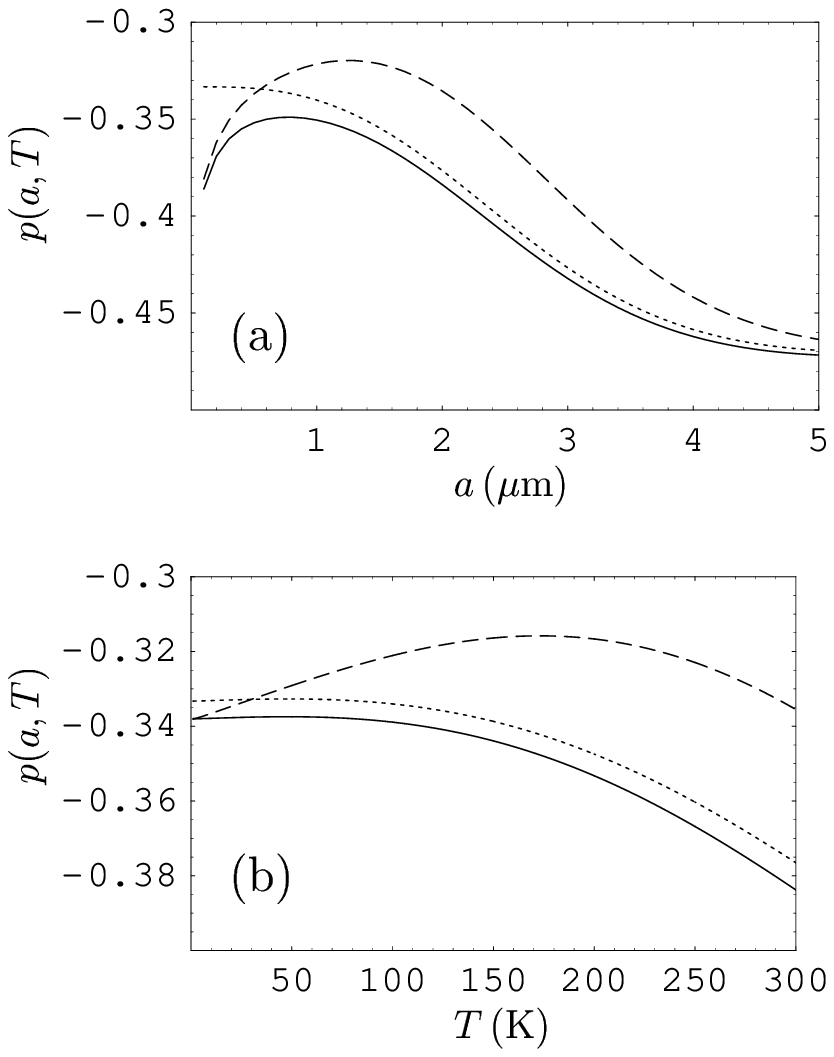}
} \vspace*{-11.cm} \caption{The quantity $p$ characterizing differences
between the two formulations of the PFA for the gradient
of the Casimir force as a function of (a) separation
at room temperature $T=300\,$K and (b) temperature at a separation
$a=2\,\mu$m. The dotted, dashed and solid lines show the results
computed using ideal metals, Drude and plasma model approaches,
respectively. The sphere radius $R=100\,\mu$m.}
\end{figure*}
\begin{figure*}[h]
\vspace*{-8.cm}
\centerline{\hspace*{2cm}
\includegraphics{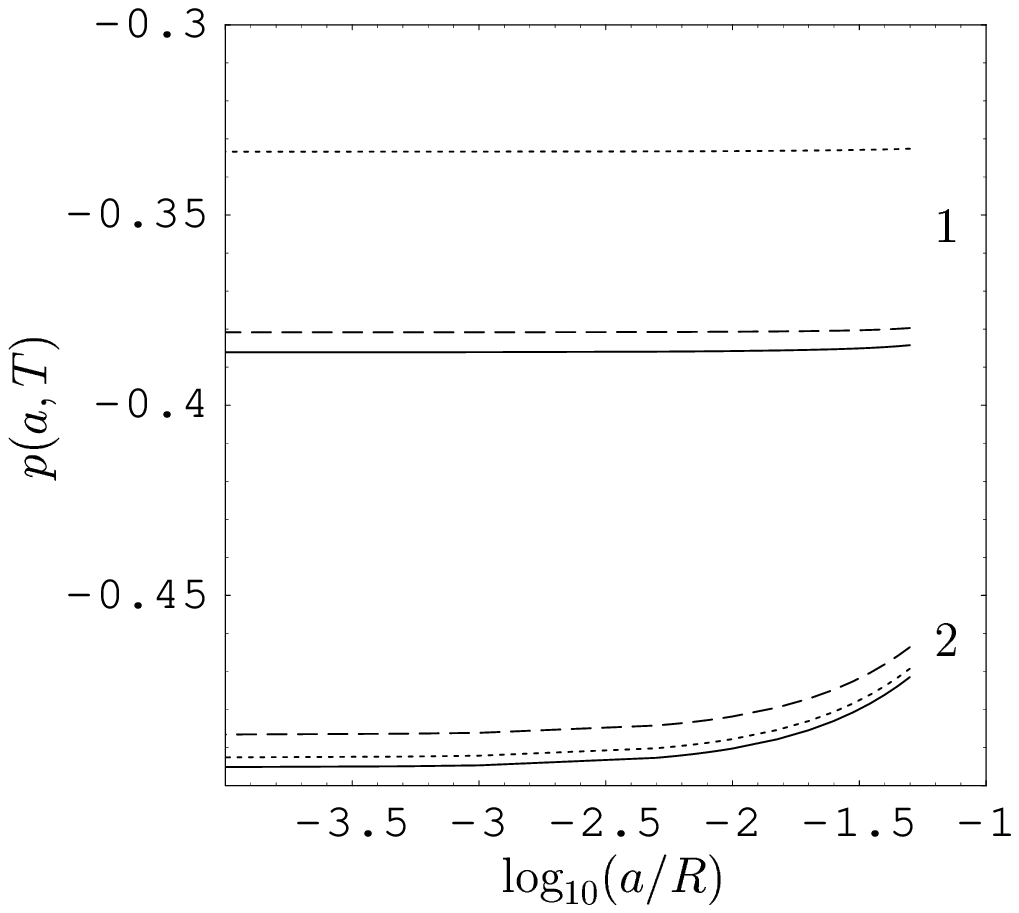}
} \vspace*{-6.5cm} \caption{The quantity $p$ characterizing differences
between the two formulations of the PFA for the gradient
of the Casimir force as a function of $a/R$ at
room temperature $T=300\,$K.
The dotted, dashed and solid lines show the results
computed using ideal metals, Drude and plasma model approaches,
respectively.
The groups of lines numerated 1 and  2 are for the respective
fixed separations $a=0.1$ and $5\,\mu$m.}
\end{figure*}
\begin{figure*}[h]
\vspace*{-4.cm}
\centerline{
\includegraphics{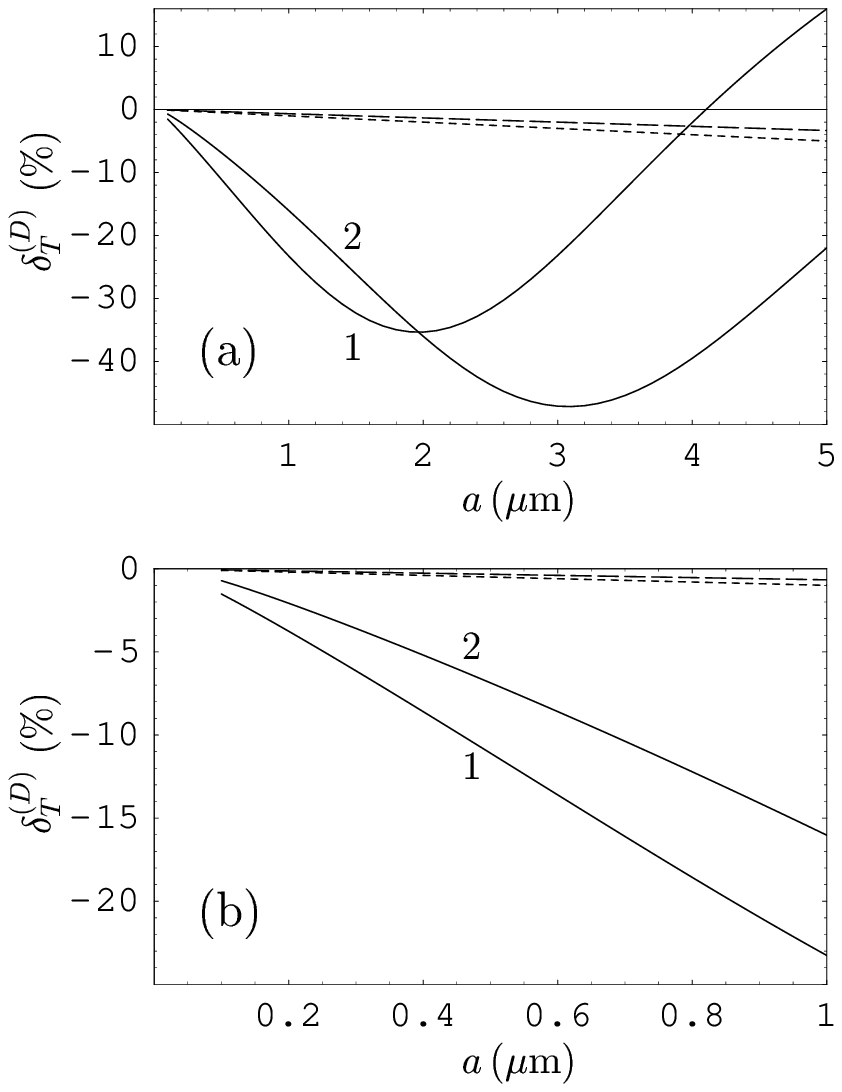}
} \vspace*{-11.cm} \caption{Relative thermal correction to the Casimir
force (the solid line 1) and to the Casimir pressure
(the solid line 2) computed using the Drude model approach at $T=300\,$K
over the separation region  (a) from 0.1 to $5\,\mu$m and
(b) from 0.1 to $1\,\mu$m. The short-dashed and long-dashed lines
show the quantity $a/R$ for $R=100$ and $150\,\mu$m, respectively. }
\end{figure*}
\begin{figure*}[h]
\vspace*{-8.cm}
\centerline{\hspace*{2cm}
\includegraphics{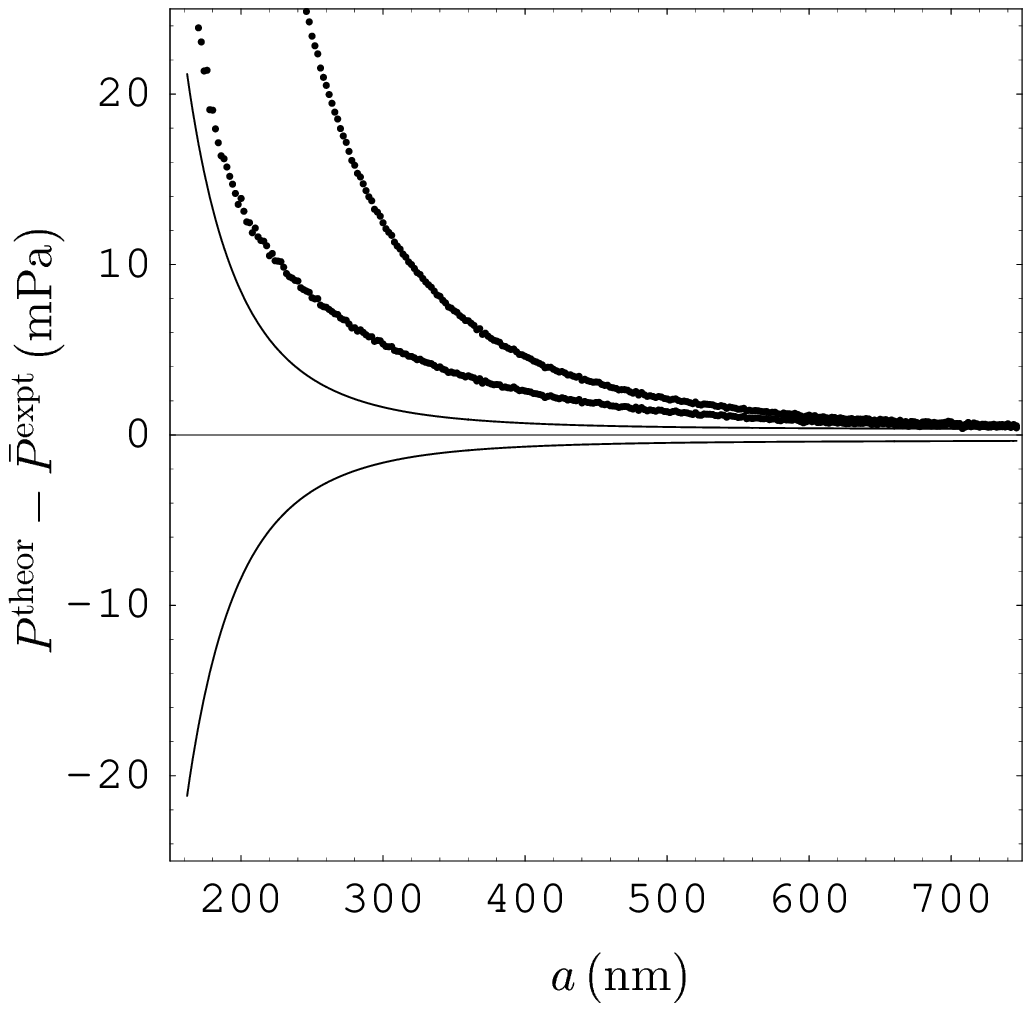}
} \vspace*{-6.5cm} \caption{Differences between the theoretical Casimir
pressures, computed by means of the Drude model approach, and mean
experimental Casimir pressures at different separations are shown as
dots. The lower and upper sets of dots correspond to the use of
conventional and alternative Drude parameters (see text for further
details). The solid lines indicate the borders of the confidence intervals
found at a 95\% confidence level.}
\end{figure*}
\begin{figure*}[h]
\vspace*{-3.cm}
\centerline{
\includegraphics{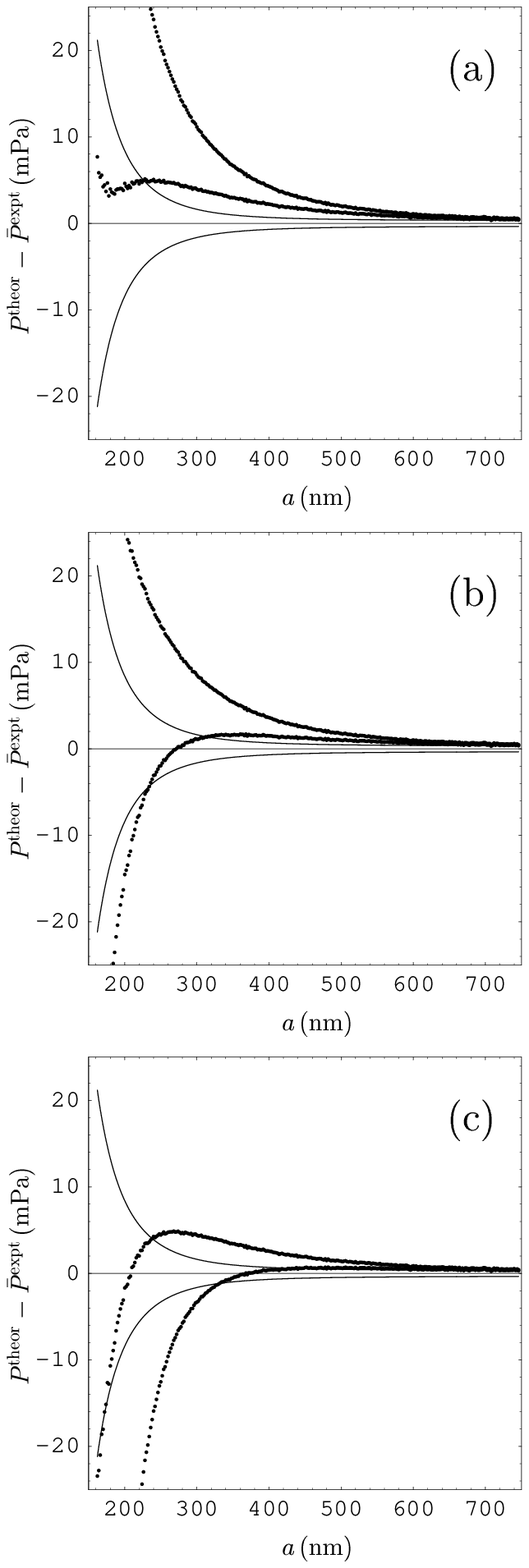}
} \vspace*{-7.5cm} \caption{Differences between the theoretical Casimir
pressures, computed by means of the Drude model approach, and mean
experimental Casimir pressures at different separations are shown as
dots. Compared to Fig.~6, all separation distances are decreased by
(a) $\Delta a=1\,$nm, (b) $\Delta a=3\,$nm and
(c) $\Delta a=6\,$nm.
}
\end{figure*}
\begin{figure*}[h]
\vspace*{-5.cm}
\centerline{\hspace*{2cm}
\includegraphics{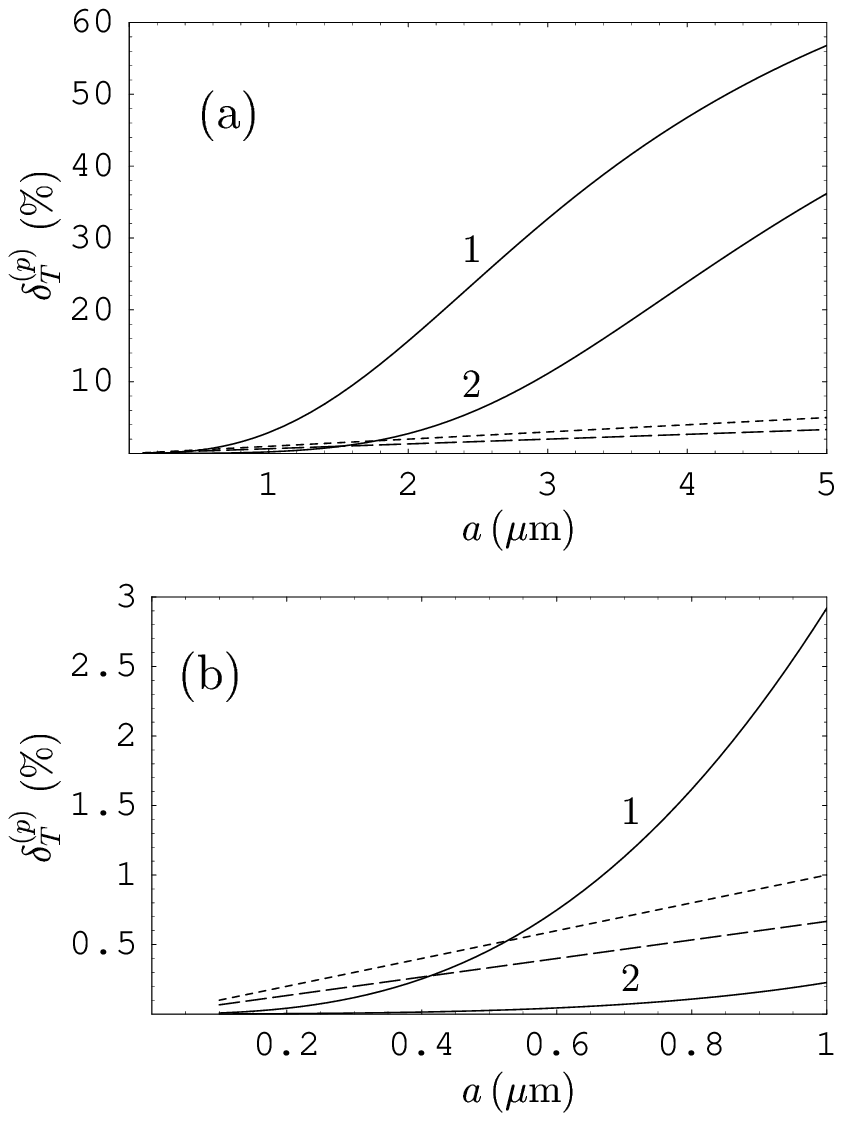}
} \vspace*{-11.cm} \caption{Relative thermal correction to the Casimir
force (the solid line 1) and to the Casimir pressure
(the solid line 2) computed using the plasma model approach at $T=300\,$K
over the separation region (a) from 0.1 to $5\,\mu$m and
(b) from 0.1 to $1\,\mu$m. The short-dashed and long-dashed lines
show the quantity $a/R$ for $R=100$ and $150\,\mu$m, respectively.}
\end{figure*}
\begin{figure*}[h]
\vspace*{-12.cm}
\centerline{\hspace*{2cm}
\includegraphics{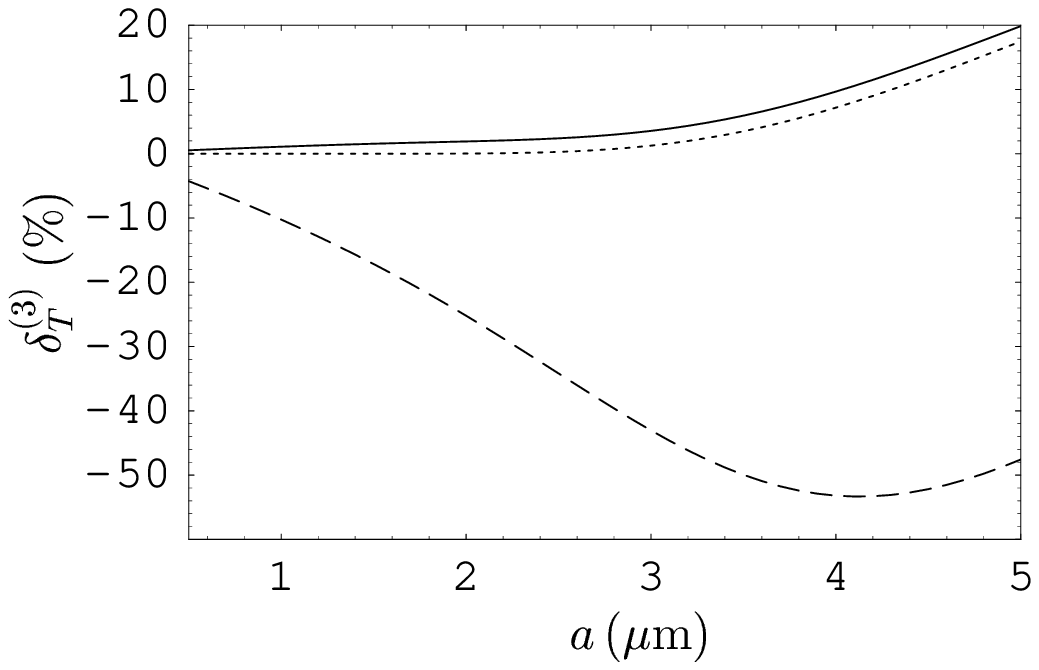}
} \vspace*{-6.cm} \caption{Relative thermal correction
to the gradient of the Casimir pressure between two parallel plates
as a function of separation computed at $T=300\,$K using ideal metals
(the dotted line), and the Drude and plasma model approaches
(the dashed and solid lines, respectively).}
\end{figure*}
\end{document}